\documentclass[prd,aps,nofootinbib]{revtex4}
\usepackage{amssymb,epsfig}

\newcommand{\Real}{\mathbb{R}}

\begin{document}
%%%%%%%%%%%%%%%%%%%%%%%%%%%%%%%

\title{Do unbounded bubbles ultimately become fenced inside a black hole?}
\author{F. S. Guzm\'an$^{1}$, L. Lehner$^{2}$ and O. Sarbach$^{1}$}

\affiliation{
${}^{1}$Instituto de F\'{\i}sica y Matem\'{a}ticas,
        Universidad Michoacana de San Nicol\'as de Hidalgo. Edificio C-3,
        Cd. Universitaria,
        C. P. 58040 Morelia, Michoac\'{a}n, M\'exico.\\
${}^{2}$Department of Physics and Astronomy, Louisiana State University, 
202 Nicholson Hall, Baton Rouge, Louisiana 70803-4001, USA}

\begin{abstract}
We examine the dynamical behavior of recently introduced bubbles in
asymptotically flat, five-dimensional spacetimes. Using numerical
methods, we find that even bubbles that initially start expanding
eventually collapse to a Schwarzschild-Tangherlini black hole.
\end{abstract}

%\pacs{11.25.-w, 11.25.Sq, 04.25.Dm}

\maketitle

%%%%%%%%%%%%%%%%%%%%%%%%%%%%%%%%%%%%%%%%%%%%%%%%%%%%%%%%%%%%%%
\section{Introduction}
\label{Sect:Intro}
%%%%%%%%%%%%%%%%%%%%%%%%%%%%%%%%%%%%%%%%%%%%%%%%%%%%%%%%%%%%%%

``Bubbles of nothing'' in higher dimensional spacetimes have been the
subject of significant attention in recent years. They have played a
key role in understanding the phase space of black hole spacetimes in
Kaluza-Klein scenarios \cite{hEtHnO05,hEgH03}, have surfaced as a
possible way around the black hole information paradox \cite{gH05} and
as mediators of non-perturbative instabilities in AdS/CFT contexts
\cite{He:2007ji}, been discussed in connections with orbifold decays
in AdS spacetimes \cite{Balasubramanian:2005bg} and have been the
subject of studies of bubble-bubble collisions
\cite{gHkM02,bFgHsS07}. However, essentially all higher dimensional
bubble studies have been presented within Kaluza-Klein scenarios as no
known non-local bubble solutions outside them were available. Recent
work by Copsey \cite{kC07a,kC07b} presents the first examples of
bubbles in asymptotically flat and AdS cases. In particular,
Ref. \cite{kC07b} provides data, at a moment of time symmetry, where
surfaces of constant radius $r$ are squashed three-spheres with the
circumference of one of the circles converging to zero when $r$
approaches some fixed positive value while these surfaces are metric
three-spheres in the asymptotic region $r\to\infty$. By judiciously
choosing the freely available geometrical variables, a ``bubble of
nothing'' can be defined. Since Copsey's construction gives data at a
moment of time a complete picture of the spacetime is not
available. While several important observations can be drawn at the
hypersurface where this data is known, like the initial growth-rate of
the bubble, key issues can only be addressed by examining the full
evolution of the spacetime subject to the provided initial data. Among
the questions one would like to answer are
\begin{itemize}
\item What is the bubble's behavior in time? In particular, if a
bubble begins expanding, does it keep on expanding? If the expansion
were exponentially fast as in the case of Kaluza-Klein bubbles
\cite{oSlL04}, the spacetime might have the ingredients for realizing
a possible violation of cosmic censorship. This would occur if the
bubble expanded so as to meet (a portion of) future null infinity at a
finite affine time of the generators of ${\cal I}^+$.
\item If the bubble were to reverse its expansion rate, what is the
end behavior?  Does the bubble collapse? If so, does it form a black
hole?
\item If a black hole is formed, is the bubble size when the horizon
forms already at the string/Planck scales? In this case the classical
evolution certainly could not be trusted and string effects should be
taken into account to reveal the bubble's final fate.
\end{itemize}
In this work we examine the evolution of some of the bubble initial
data presented by Copsey and answer the above questions.

Our article is organized as follows. In section \ref{Sect:Idata} we
review Copsey's initial data representing exact bubble solutions at a
moment of time symmetry. The future of this data is then obtained by
solving the Cauchy problem by numerical methods. The system of
evolution and constraint equations, the coordinate choices and
boundary conditions used to evolve this initial data set as well as
the numerical implementation are discussed in section
\ref{Sect:Evolution}. In section \ref{Sect:Results} we present the
results of the evolutions carried out and we end in section
\ref{Sect:Conclusion} with a brief discussion and conclusions drawn
from this work. Technicalities like the computation of the curvature
tensor and curvature invariants and a summary of the
Schwarzschild-Tangherlini solution can be found in the appendices.

%%%%%%%%%%%%%%%%%%%%%%%%%%%%%%%%%%%%%%%%%%%%%%%%%%%%%%%%%%%%%%
\section{Prototype metric and initial data}
\label{Sect:Idata}
%%%%%%%%%%%%%%%%%%%%%%%%%%%%%%%%%%%%%%%%%%%%%%%%%%%%%%%%%%%%%%

Recently, Copsey \cite{kC07b} introduced bubbles which are outside the
traditionally Kaluza-Klein framework. These bubbles, for the
asymptotically flat case, and at a moment of time-symmetry, are
described by a $t = const$ slice of a metric of the form
\begin{equation}
ds^2 = -dt^2 + \frac{dr^2}{W(r)} + \frac{r^2}{4}\left[ 
   H(r) \left( dz + \cos\vartheta\, d\varphi \right)^2
 + d\vartheta^2 + \sin^2\vartheta\, d\varphi^2 \right], 
\label{Eq:PrototypeMetric}
\end{equation}
where $W$ and $H$ are smooth functions of $r$ which converge to $1$ as
$r\to\infty$ and where $(\vartheta,\varphi,z)\in [0,\pi) \times
[0,2\pi) \times [0,4\pi)$ are Euler angles on the three-sphere
$S^3$. If $W\equiv H\equiv 1$, this metric describes five-dimensional
Minkowski space. This can be seen by introducing the coordinates
\begin{displaymath}
x + i y = r\, e^{\frac{i}{2}(z-\varphi)}\sin\left(\frac{\vartheta}{2}\right),
\qquad
X + i Y = r\, e^{\frac{i}{2}(z+\varphi)}\cos\left(\frac{\vartheta}{2}\right),
\end{displaymath}
in terms of which the metric (\ref{Eq:PrototypeMetric}) with $W\equiv
H\equiv 1$ assumes the form
\begin{displaymath}
ds^2 = -dt^2 + dx^2 + dy^2 + dX^2 + dY^2.
\end{displaymath}
More generally, using
\begin{displaymath}
dr = n_i dx^i, \qquad
dz + \cos\vartheta\, d\varphi = m_i dx^i,
\end{displaymath}
where $(x^i) = (x,y,X,Y)$, $(n_i) = (x,y,X,Y)/r$ and $(m_i) = (-y,x,-Y,X)/r$,
we find that
\begin{displaymath}
ds^2 = -dt^2 + h_{ij} dx^i dx^j
     = -dt^2 + \left[ 
  \delta_{ij} + \frac{1 - W}{W}\, n_i n_j + (H-1) m_i m_j \right] dx^i dx^j.
\end{displaymath}
Therefore, the metric (\ref{Eq:PrototypeMetric}) is asymptotically
flat if the functions $W-1$ and $H-1$ decay to zero fast enough as
$r\to\infty$. In particular, if $W = 1 - C_1/r^2 + O(r^{-3})$, $H = 1
- C_2/r^2 + O(r^{-3})$, $H' = 2C_2/r^3 + O(r^{-4})$ we obtain a finite
ADM mass
\begin{displaymath}
M_{ADM} = \frac{1}{16\pi} \lim\limits_{r\to\infty} \int\limits_{S_r^3}
 \sum\limits_{i,j=1}^4
 \left( \partial_i h_{ij} - \partial_j h_{ii} \right) dS_j
 = \frac{\pi}{8} (3C_1 - C_2),
\end{displaymath}
where $S_r^3$ denotes the three-sphere with radius $r$ and $dS_j = n_j
dS$ is the area element on $S_r^3$.

Copsey's construction also assumes that the functions $W$ and $H$ both
have a single root at some $r_0 > 0$ and are both strictly positive
for $r > r_0$. This means that the circumference of the circles
determined by the orbits of the Killing field $\partial_z$ shrinks to
zero at $r = r_0$. In order to understand the geometry near $r=r_0 $,
we replace $r$ by the new radial coordinate
\begin{displaymath}
R = \sqrt{r^2 - r_0^2}\, , \qquad r \ge r_0\, .
\end{displaymath}
Then, the metric (\ref{Eq:PrototypeMetric}) can be rewritten as
\begin{equation}
ds^2 = -dt^2 + \frac{R^2 dR^2}{(R^2 + r_0^2) W} + \frac{R^2 + r_0^2}{4}\left[ 
   H\left( dz + \cos\vartheta\, d\varphi \right)^2
 + d\vartheta^2 + \sin^2\vartheta\, d\varphi^2 \right].
\label{Eq:PrototypeMetricNewCoords}
\end{equation}
Because of the assumptions on $W$ and $H$, they have the form
\begin{eqnarray}
W &=& 2\alpha^2\;\frac{r - r_0}{r_0} + O\left( \frac{r-r_0}{r_0} \right)^2
   =  \alpha^2\left( \frac{R}{r_0} \right)^2 + O\left( \frac{R}{r_0} \right)^4,
\nonumber\\
H &=&  2\beta^2\;\frac{r - r_0}{r_0} + O\left( \frac{r-r_0}{r_0} \right)^2
   =  \beta^2\left( \frac{R}{r_0} \right)^2 + O\left( \frac{R}{r_0} \right)^4,
\nonumber
\end{eqnarray}
near $R = 0$, where $\alpha$ and $\beta$ are two strictly positive
constants. Therefore, as $R$ tends to zero, the metric has the form
\begin{displaymath}
ds^2 \simeq -dt^2 + \frac{dR^2}{\alpha^2} + \frac{\beta^2}{4} R^2 
   \left( dz + \cos\vartheta\, d\varphi \right)^2
 + \frac{r_0^2}{4} \left[ d\vartheta^2 + \sin^2\vartheta\, d\varphi^2 \right]. 
\end{displaymath}
Since $z$ has period $4\pi$, there is a conical singularity at $R=0$
unless $\alpha\beta = 1$. If $\alpha\beta=1$, we may replace $R$ and
$z$ with the Cartesian coordinates $(u,v)$, which are defined by
\begin{displaymath}
u + i v = \frac{R}{\sqrt{\alpha}}\, e^{i z/2}.
\end{displaymath}
In terms of these, the metric (\ref{Eq:PrototypeMetricNewCoords})
assumes the form
\begin{eqnarray}
ds^2 &=& -dt^2 + du^2 + dv^2 + \frac{r_0^2 + \alpha(u^2 + v^2)}{4}
   \left[ d\vartheta^2 + \sin^2\vartheta\, d\varphi^2 \right]
\nonumber\\
 &+& f(u^2+v^2) (u\, du + v\, dv)^2 + h(u^2+v^2) \left[ 
      2(u\, dv - v\, du) + (u^2+v^2)\cos\vartheta\, d\varphi \right]^2,
\nonumber
\end{eqnarray}
where $f$ and $h$ are smooth functions of $u^2 + v^2$. Therefore, if
$\alpha\beta=1$, we obtain a smooth, regular asymptotically flat
manifold with topology $\Real_t \times \Real^2 \times S^2$. At each
fixed time $t$, the bubble is determined by the two-sphere at which
the Killing field $\partial_z$ vanishes.

\subsection{Initial data}

Data at the moment of time symmetry automatically solves the momentum
constraint and one is left with having to satisfy the Hamiltonian
constraint only. For a $t=const$ section of a metric of the form
(\ref{Eq:PrototypeMetric}), the Hamiltonian constraint in vacuum
yields
\begin{equation}
\left[ \frac{H''}{H} +\frac{4}{r}\frac{H'}{H} - \frac{1}{2}\frac{H'^2}{H^2} 
 + \frac{6}{r^2} \right] W 
 + \left[ \frac{1}{2}\frac{H'}{H} + \frac{3}{r} \right] W'
 + \frac{2}{r^2}\left( H - 4 \right) = 0,
\label{Eq:TimeSymmetricHam}
\end{equation}
where a prime denotes differentiation with respect to $r$. A
convenient way to solve this equation is to freely specify $H$ and
integrate the resulting linear equation for $W$. In order to analyze
this, let $H$ be an arbitrary smooth function with the following
properties: $H$ has a single root at some point $r=r_0 > 0$, $H(r) > 0
$ for all $r > r_0$ and $H(r) = 1 - C_2/r^2 + O(r^{-3})$, $H'(r) =
2C_2/r^3 + O(r^{-4})$ and $H''(r) = -6C_2/r^4 + O(r^{-5})$ for $r \to
\infty$. Furthermore, we assume that the expansion along outgoing null
radial geodesics, which is proportional to $6H + rH'$, is strictly
positive for all $r \geq r_0$. This condition ensures the initial data
does not contain any apparent horizons \cite{kC07b}.

For the following, it is convenient to replace the function $W$ by a
new function $\zeta$, defined by $W(r) = H(r)( 1 + \zeta(r))$, and to
introduce the dimensionless compactified coordinate $s := r_0/r$ which
varies from $0$ to $1$. Then, Eq. (\ref{Eq:TimeSymmetricHam}) reads
\begin{displaymath}
\frac{s}{2} (6H - s H_s)\zeta_s = (s^2 H_{ss} - 5s H_s + 6H)\zeta
 + \left[ s^2 H_{ss} - 5s H_s + 8(H-1) \right],
\end{displaymath}
where the subscript $s$ denotes differentiation with respect to $s$.
The general solution to this equation is
\begin{equation}
\zeta(s) = \zeta_0(s)\left[ C_0 + J(s) \right],
\label{Eq:zetaSolution}
\end{equation}
where $C_0$ is a constant and
\begin{displaymath}
\zeta_0(s) = \frac{s^2}{(6H - s H_s)^2}\; e^{2I(s)}, \qquad
J(s) = 2\int\limits_0^s (6H - \tau H_\tau) 
 \frac{\tau^2 H_{\tau\tau} - 5\tau H_\tau + 8(H-1)}{\tau^3}\, e^{-2I(\tau)} 
d\tau,
\end{displaymath}
with
\begin{displaymath}
I(s) = \int\limits_0^s \frac{H_\tau d\tau}{6H - \tau H_\tau}\; .
\end{displaymath}
Because of the above assumptions on $H$, these integrals are
well-defined and converge to a finite value as $s\to 1$. In
particular, notice that the asymptotic behavior on $H$ implies that
$\tau^2 H_{\tau\tau} - 5\tau H_\tau + 8(H-1) = O(\tau^3)$ so that the
integrand in the expression for $J(s)$ is well-defined at
$\tau=0$. Furthermore, we see that $\zeta(s) = O(s^2)$ for small $s >
0$, hence the function $W$ has the same asymptotic behavior as $H$ and
possesses a root at $r=r_0$. The constant $C_0$ is determined from the
requirement of the absence of a conical singularity at $r=r_0$,
$H_s(1) W_s(1) = 4$, which yields
\begin{displaymath}
C_0 = \left[ 4 - H_s(1)^2 \right] e^{-2I(1)} - J(1).
\end{displaymath}
Finally, we notice that without further restriction on $H$ there is no
guarantee that $W$ is positive for all $r > r_0$. However, the
condition $H < 4$ is sufficient to guarantee positivity of $W$ for $r
> r_0$. Indeed, Eq. (\ref{Eq:TimeSymmetricHam}) and the condition $6H
+ r H' > 0$ then imply that $W'(r_1) > 0$ if $r_1$ is a zero of $W$
which shows that the sign of $W$ cannot change from positive to
negative when $r$ increases.

In order to understand the initial dynamics of the bubble, we compute
the initial acceleration of the bubble area ${\cal A}$ with respect to
proper time $\tau$ at the bubble. In terms of the general metric
(\ref{Eq:MetricGeneral}) the bubble's area and proper time are given
by
\begin{displaymath}
{\cal A} = \left. 4\pi e^{2c} \right|_{r=r_0}\; ,\qquad
d\tau = \left. e^d dt \right|_{r=r_0},
\end{displaymath}
respectively. Taking the second derivative with respect to proper
time, using the evolution equation $R_{33} = 0$ in appendix
\ref{App:Exp} and the initial values
\begin{displaymath}
R = r, \qquad
a = -\frac{1}{2}\log(W), \qquad
b = \log\left( \frac{r}{2} \right) + \frac{1}{2}\log(H), \qquad
c = \log\left( \frac{r}{2} \right),
\end{displaymath}
$\dot{a}=\dot{b}=\dot{c}=0$ in (\ref{Eq:MetricGeneral}), we obtain
\begin{equation}
\frac{d^2 {\cal A}}{d\tau^2} = -2\pi W_s(1) - 8\pi
 = -8\pi\left[ 1 + \frac{1}{H_s(1)} \right].
\end{equation}
Therefore, the bubble starts expanding if and only if $-1 < H_s(1) <
0$. Finally, the ADM mass is
\begin{displaymath}
M_{ADM} = \frac{\pi r_0^2}{8}
\left( 2\frac{C_2}{r_0^2} + \frac{C_0}{36} \right).
\end{displaymath}

Some examples of solutions, and the ones we will employ in our
simulations are
\begin{enumerate}
\item The choice $W = H$ simplifies Eq. (\ref{Eq:TimeSymmetricHam}) to
the following linear equation for $H$,
\begin{displaymath}
H'' + \frac{7}{r} H' + \frac{8}{r^2}(H - 1) = 0,
\end{displaymath}
which has the general solution
\begin{equation}
H(r) 
 = 1 - a_0\left( \frac{r_0}{r} \right)^2 - a_1\left( \frac{r_0}{r} \right)^4
\label{Eq:H}
\end{equation}
with $a_0 + a_1 = 1$. Notice that $H(r_0) = 0$ and that $H$ can also
be written in the form
\begin{displaymath}
H(r) = \left[ 1 - \left( \frac{r_0}{r} \right)^2 \right]
\left[ 1 + a_1\left( \frac{r_0}{r} \right)^2 \right]
\end{displaymath}
from which we see that a necessary and sufficient condition for $H$ to
have a single root at $r_0$ and to be strictly positive for $r > r_0$
is that $a_1 > -1$, or, equivalently, $a_0 < 2$. Furthermore, $H$
converges to one as $r\to\infty$. In order to verify the regularity
condition at $r=r_0$, we compute $2\alpha^2 = r_0 H'(r_0) = 2(1+a_1)$.
Hence, $\alpha^2 = \beta^2 = 1 + a_1$ and the regularity condition
$\alpha\beta=1$ forces the choice $a_0=1$, $a_1 = 0$.
\item
More general initial data can be obtained by employing the function
$H$ given in Eq. (\ref{Eq:H}) but relaxing the condition $W=H$. In
this case, the integrals in Eq. (\ref{Eq:zetaSolution}) can be
performed analytically \cite{kC07b} with the result
\begin{equation}
\zeta(r) = c_1 \left( \frac{r_0}{r} \right)^2
\left| \frac{ a_1\left( \frac{r_0}{r} \right)^2 + a_0 - b_0 }{a_1 + a_0 - b_0}
\right|^{-\frac{a_0}{2b_0}-1}
\left| \frac{ a_1\left( \frac{r_0}{r} \right)^2 + a_0 + b_0 }{a_1 + a_0 + b_0}
\right|^{\frac{a_0}{2b_0}-1},
\label{Eq:zeta}
\end{equation}
with
\begin{displaymath}
b_0 = \sqrt{a_0^2 + 3a_1}, \qquad
c_1 = -\frac{a_1(a_1+2)}{(1+a_1)^2}\, .
\end{displaymath}
If $a_1 = 0$, $\zeta\equiv 0$ and we recover the previous solution
with $W=H$. However, the more general family of solutions given by $H$
as in Eq. (\ref{Eq:H}) and $W = H(1 + \zeta)$ with $\zeta$ given in
Eqs. (\ref{Eq:zeta}) allows for initially collapsing and expanding
bubbles \cite{kC07b} since the initial acceleration is
\begin{equation}
\frac{d^2 {\cal A}}{d\tau^2} = 4\pi\,\frac{2a_0 - 3}{2 - a_0}\; .
\end{equation}
Therefore, the bubble initially expands for $3/2 < a_0 < 2$ and
initially collapses for $a_0 < 3/2$. The parameter $a_0$ determines
the ADM mass through
\begin{displaymath}
M_{ADM} = \frac{\pi r_0^2}{8}\left[ 2 a_0 + 3\frac{a_1(a_1+2)}{(1+a_1)^2}
 \left| \frac{ a_0 - b_0 }{a_1 + a_0 - b_0} \right|^{-\frac{a_0}{2b_0}-1}
 \left| \frac{ a_0 + b_0 }{a_1 + a_0 + b_0} \right|^{\frac{a_0}{2b_0}-1}
 \right].
\end{displaymath}

\item
A simple and interesting example is given by the choice $H(s) =
(1-s^2)^2 + \varepsilon s^2(1-s)$, where $0 < \varepsilon < 1$. This
choice satisfies all the required assumptions on $H$, and $0 < H(s) <
1 + \varepsilon$ for $0 \leq s < 1$. Furthermore, $H_s(1) =
-\varepsilon$, so one can construct bubbles with arbitrarily large
initial acceleration.

\item The following ansatz is also suggested in \cite{kC07b},
\begin{equation}
H(s) = (1-s^2)^4 + 4\varepsilon s^2(1-s)
 - \frac{4s^2 (1-s)^2 c_2}{1+\frac{c_2}{\epsilon} (1-s)}\; ,
\end{equation}
which leads to initially hyperexpanding bubbles.
\end{enumerate}

%%%%%%%%%%%%%%%%%%%%%%%%%%%%%%%%%%%%%%%%%%%%%%%%%%%%%%%%%%%%%%
\section{Evolution}
\label{Sect:Evolution}
%%%%%%%%%%%%%%%%%%%%%%%%%%%%%%%%%%%%%%%%%%%%%%%%%%%%%%%%%%%%%%

Here, we discuss our method for obtaining the time evolution of the
time-symmetric bubble configurations discussed in the previous
section. Motivated by the form (\ref{Eq:PrototypeMetricNewCoords}) of
the prototype metric, we perform the following rescaling in the
general line element (\ref{Eq:MetricGeneral})
\begin{equation}
d \mapsto d, \qquad
a \mapsto a, \qquad
b \mapsto b + \log\left( \frac{R}{2} \right), \qquad
c \mapsto c + \log\left( \frac{\sqrt{R^2 + r_0^2}}{2} \right).
\label{Eq:Rescaling}
\end{equation}
The metric now takes the form
\begin{equation}
ds^2 = -e^{2d} dt^2 + e^{2a} dR^2 + \frac{1}{4}\left[ 
         e^{2b} R^2 \left( dz + \cos\vartheta\, d\varphi \right)^2 
       + e^{2c} (R^2 + r_0^2) d\Omega^2 \right], \qquad R \geq 0.
\label{Eq:MetricEvolution}
\end{equation}
The functions $a$, $b$, $c$ and $d$ which depend only on $t$ and $R$
must satisfy the following boundary conditions. As $R \to \infty$ we
require that these functions converge to zero fast enough for
quantities like the ADM mass to be defined. Near $R=0$ we require that
these functions are smooth and satisfy the conditions $a'(t,0) =
b'(t,0) = c'(t,0) = d'(t,0) = 0$. Furthermore, we impose the condition
$a(t,0) - b(t,0) = 0$ which ensures that there is no conical
singularity at $R=0$.

\subsection{Gauge conditions}

We find it convenient to impose the following family of gauge
conditions on the logarithm of the lapse $d$,
\begin{equation}
d = a + \lambda(b + 2c),
\label{Eq:GaugeCondition}
\end{equation}
where $\lambda$ is a fixed parameter. For $\lambda=0$, this condition
implies that the two-metric $-e^{2d} dt^2 + e^{2a} dR^2$ is in the
conformal flat gauge. As we will see, the principal part of the
evolution equations is governed by the d'Alembertian with respect to
this metric. Since the two-dimensional d'Alembertian operator is
conformally covariant, the resulting evolution equations are
semi-linear. In particular, this implies that the characteristic
speeds do not depend on the solution that is being evolved, so there
cannot be shock formation due to the crossing of characteristics. When
$\lambda=1$, the gauge condition (\ref{Eq:GaugeCondition}) is strongly
related to the densitized lapse condition often encountered in
hyperbolic formulations of Einstein's field equations. Indeed, the
square root of the determinant of the four metric belonging to
(\ref{Eq:MetricEvolution}) is $e^{a + b + 2c} R(R^2 +
r_0^2)\sin\vartheta/8$, so (\ref{Eq:GaugeCondition}) sets the lapse
$e^d$ equal to the square root of the determinant of the four metric
divided by the factor $R(R^2 + r_0^2)\sin\vartheta/8$ which is
singular at the bubble, at the poles $\vartheta = 0,\pi$ and at
$R=\infty$. Since this gauge condition is essentially the time
harmonic gauge condition integrated in time, it is convenient for its
singularity avoidance behavior. In our simulations below, we use both
choices $\lambda=0$ and $\lambda=1$.

\subsection{Evolution equations}

In the gauge (\ref{Eq:GaugeCondition}), the evolution equations can be
written as a coupled system of three wave equations for the fields
$a$, $b$ and $c$ which are obtained from the Einstein vacuum equations
$R_{11} - (\lambda+1) G_{00} = 0$, $R_{22} = 0$ and $R_{33} = 0$,
respectively. The resulting system has the form
\begin{displaymath}
\ddot{u} = e^{2\lambda(b+2c)}\left[ u'' + \frac{1}{R}\, M u' \right] 
 + {\cal F}(R,u,\dot{u},u'),
\end{displaymath}
where $u := (a,b,c)^T$, $M$ is the constant $3\times 3$ matrix
\begin{displaymath}
M = \left( \begin{array}{ccc}
  \lambda & -2\lambda & -2(\lambda+1) \\
        0 & \lambda+2 & 2(\lambda+1) \\
        0 &         0 & 1
 \end{array}\right),
\end{displaymath}
and ${\cal F}$ is a nonlinear function of $R$, $u$, $\dot{u}$ and $u'$
which is regular at $R=0$. Provided that $u$ is smooth enough, the
evolution equations are regular at $R=0$ since then the boundary
condition $u'(t,0) = 0$ implies that
\begin{displaymath}
\lim\limits_{R\to 0} \frac{1}{R}\, M u' = \left. M u'' \right|_{R=0}\; .
\end{displaymath}
For the following, we find it convenient to replace $a$, $b$ and $c$
by the linear combinations $A = a + \lambda b + 2(\lambda+1)c$, $B = b
+ 2c$ and $C = c$ which diagonalize $M$. Using the expressions in the
appendix and taking into account the rescaling (\ref{Eq:Rescaling}) we
obtain
\begin{eqnarray}
\ddot{A} &=& \frac{1}{R^{\lambda}}
   \left[ R^{\lambda} e^{2\lambda B} A' \right]'
 + e^{2\lambda B}\left[ -3(\lambda-1)C'^2 
 + \frac{2R}{r_0^2 + R^2}
   \left( \lambda A' + 2\lambda B' - 3(\lambda - 1)C' \right)
 + \frac{2(1+\lambda)r_0^2 + (3\lambda+1)R^2}{(r_0^2 + R^2)^2} \right]
\nonumber\\
 &+& 2\lambda\dot{A}\dot{B} - \lambda(\lambda+1)\dot{B}^2 
  - 3(\lambda+1)\dot{C}^2 + (\lambda+3) V_0 - (\lambda+1) V_1\; ,
\label{Eq:A}\\
\ddot{B} &=& \frac{1}{R^{\lambda+2}}\, e^{(\lambda-1)B}
   \left[ R^{\lambda+2} e^{(\lambda+1)B} B' \right]'
 + 2\frac{e^{2\lambda B}}{r_0^2 + R^2}
   \left[ (\lambda+2)R B' + 3 \right] + (\lambda-1)\dot{B}^2 + 2V_0 - 2V_1\; ,
\label{Eq:B}\\
\ddot{C} &=& \frac{1}{R}\, e^{(\lambda-1)B}
   \left[ R e^{(\lambda+1)B} C' \right]'
 + \frac{e^{2\lambda B}}{r_0^2 + R^2}
   \left[ (\lambda+1)R B' + 2R C' + 2 \right]
 + (\lambda-1)\dot{B} \dot{C} + 2V_0 - V_1\; ,
\label{Eq:C}
\end{eqnarray}
where $V_0 = R^2(r_0^2 + R^2)^{-2} e^{2(A+B-6C)}$ and $V_1 = 4(r_0^2 +
R^2)^{-1} e^{2(A-3C)}$. These equations are regular at $R=0$ provided
the boundary conditions $A'=B'=C'=0$ at $R=0$ are satisfied and the
fields are smooth enough. Eqs. (\ref{Eq:A},\ref{Eq:B},\ref{Eq:C}) are
subject to the Hamiltonian constraint ${\cal H} = 0$ and to the
momentum constraint ${\cal M} = 0$, where
\begin{eqnarray}
{\cal H} &\equiv& -e^{2d} G_{00} 
\nonumber\\
 &=& \frac{e^{(\lambda-1)B}}{R^{\lambda+2}}
     \left[ R^{\lambda+2} e^{(\lambda+1)B} B' \right]'
 - e^{2\lambda B}\left[ \frac{1}{R}\, A' + A' B' - 3C'^2
    + \frac{2R}{r_0^2+R^2}\left( A' - (\lambda+1)B' - 3C' \right)
    - \frac{4 r_0^2 + 3 R^2}{(r_0^2 + R^2)^2} \right]
\nonumber\\
 &-& \dot{A}\dot{B} + \lambda\dot{B}^2 + 3\dot{C}^2 + V_0 - V_1\; ,
\\
{\cal M} &\equiv& -e^{2d} R_{01}
\nonumber\\
 &=& e^{\lambda B}\left[ \dot{B}' - \dot{A} B' 
  - \dot{B}\left( A' - (\lambda+1)B' \right) + 6\dot{C} C'
  - \frac{1}{R}\left( \dot{A} - (\lambda+1)\dot{B} \right)
  - \frac{2R}{r_0^2 + R^2}\left( \dot{A} - \lambda\dot{B} - 3\dot{C} \right)
 \right].
\end{eqnarray}
Notice that
\begin{equation}
\lim\limits_{R\to 0} R{\cal M} = -\left. e^{\lambda B}
  \left( \dot{A} - (\lambda+1)\dot{B} \right) \right|_{R=0}\; ,
\end{equation}
hence the satisfaction of the momentum constraint implies the
condition $a(t,0) - b(t,0) = A(t,0) - (\lambda+1)B(t,0) = 0$ for the
avoidance of the conical singularity at $R=0$. In the next subsection,
we show that the evolution equations and the regularity conditions
$A'(t,0)=B'(t,0)=C'(t,0)=0$ imply that the constraints ${\cal H} =
{\cal M} = 0$ are satisfied everywhere and at each time if satisfied
initially.

\subsection{Propagation of the constraints}

As a consequence of the twice contracted Bianchi identities (see
appendix \ref{App:Exp}) and the evolution equations which imply
$G_{11} = G_{00}$, $G_{22} = G_{33} = G_{44} = -\lambda G_{00}$ we
find that the constraint variables ${\cal H}$ and ${\cal M}$ obey the
following linear evolution system
\begin{eqnarray}
\dot{\cal H} &=& (3\lambda - 1)\dot{B}\, {\cal H}
 + e^{\lambda B}\left[ {\cal M}' 
 + \left( B' + \frac{r_0^2 + 3 R^2}{R(r_0^2 + R^2)} \right){\cal M} \right],\\
\dot{\cal M} &=& (2\lambda - 1)\dot{B}\, {\cal M}
 + e^{\lambda B}\left[ {\cal H}' + (1 + \lambda)
   \left( B' + \frac{r_0^2 + 3 R^2}{R(r_0^2 + R^2)} \right){\cal H} \right].
\end{eqnarray}
This system can be simplified by introducing the rescaled variables
\begin{displaymath}
\tilde{\cal H} := R(r_0^2 + R^2) e^B {\cal H}, \qquad
\tilde{\cal M} := R(r_0^2 + R^2) e^B {\cal M}.
\end{displaymath}
In terms of these, the constraint propagation system reads
\begin{eqnarray}
\dot{\tilde{\cal H}} &=& 3\lambda \dot{B}\, \tilde{\cal H}
 + e^{\lambda B}\tilde{\cal M}', 
\label{Eq:CProp1}\\
\dot{\tilde{\cal M}} &=& 2\lambda \dot{B}\, \tilde{\cal M}
 + e^{\lambda B}\left[ \tilde{\cal H}' + \lambda\left(
     B' + \frac{r_0^2 + 3 R^2}{R(r_0^2 + R^2)} \right) \tilde{\cal H} \right].
\label{Eq:CProp2}
\end{eqnarray}
Notice that the regularity conditions $A'(t,0) = B'(t,0) = C'(t,0) =
0$ at $R=0$ imply that $\lim\limits_{R\to 0} \tilde{\cal H} = 0$ and
that $\lim\limits_{R\to 0} \tilde{\cal M}$ exists. Therefore,
Eqs. (\ref{Eq:CProp1},\ref{Eq:CProp2}) are regular at $R=0$. Defining
the following ``energy'' norm for the constraint variables
\begin{displaymath}
{\cal E}(t) := \frac{1}{2} \int\limits_0^\infty 
 \left( \tilde{\cal H}^2 + \tilde{\cal M}^2 \right) 
 R^\lambda (r_0^2 + R^2)^\lambda dR,
\end{displaymath}
taking a time derivative and using
Eqs. (\ref{Eq:CProp1},\ref{Eq:CProp2}) we obtain
\begin{equation}
\frac{d}{dt} {\cal E} = 
 \left. R^\lambda (r_0^2 + R^2)^\lambda e^{\lambda B}\, 
 \tilde{\cal H}\, \tilde{\cal M} \right|_{R=0}^\infty
 + \lambda \int\limits_0^\infty \dot{B}
   \left( 3\tilde{\cal H}^2 + 2\tilde{\cal M}^2 \right)
    R^\lambda (r_0^2 + R^2)^\lambda dR.
\label{Eq:ConstraintEstimate}
\end{equation}
The boundary term on the right-hand side vanishes because of the
regularity conditions at $R=0$ and under the assumption that all
fields fall off sufficiently fast as $R \to \infty$. If the quantity
$\dot{B}$ is bounded, we can estimate the integral term on the
right-hand side by a constant $K$ times the energy norm ${\cal E}$,
and we obtain an estimate of the form
\begin{displaymath}
{\cal E}(t) \leq e^{|\lambda| K t} {\cal E}(0).
\end{displaymath}
This shows that if the constraints are satisfied initially, they are
also satisfied for all $t > 0$ for which a smooth enough solution to
the evolution equations exists. More generally, this inequality shows
that convergence of ${\cal E}(0)$ to zero implies that ${\cal E}(t)$
also converges to zero for each such $t$. This is important for the
numerical implementation below where the initial data does not satisfy
the constraints exactly due to truncation errors, but where the
initial constraint violation converges to zero with increasing
resolution. Finally, we notice that in the particular gauge where
$\lambda=0$ the ``energy'' norm ${\cal E}$ cannot grow in time.

\subsection{Outer boundary conditions}

In our numerical implementation below, we truncate the domain at some
large radius $R_{max} \gg r_0$ and impose the following boundary
conditions at $R = R_{max}$. First, we enforce the momentum constraint
$\left. {\cal M} \right|_{R=R_{max}} = 0$, which yields the boundary
condition
\begin{equation}
\left. \dot{B}' \right|_{R=R_{max}}
 = \left[ \dot{A} B' + \dot{B} ( A' - (\lambda+1)B' ) - 6\dot{C} C'
 + \frac{1}{R}\left( \dot{A} - (\lambda+1)\dot{B} \right)
 + \frac{2R}{r_0^2+R^2}\left( \dot{A} - \lambda\dot{B} - 3\dot{C} \right)
 \right]_{R=R_{max}}\; .
\label{Eq:CPBC}
\end{equation}
When estimating the constraint errors, $\infty$ has to be replaced by
$R_{max}$ in the expression for ${\cal E}(t)$ above. One then obtains
the same equality as in Eq. (\ref{Eq:ConstraintEstimate}) but with
$\infty$ replaced by $R_{max}$ on the right-hand side of that
equation. The enforcing of the momentum constraint then guarantees
that the boundary term vanishes. Therefore, we conclude as before that
the constraint errors converge to zero at each fixed $t > 0$ if the
initial constraint errors converge to zero.

Finally, we impose a Sommerfeld-like boundary condition on the fields
$A$ and $C$,
\begin{equation}
\left. \dot{A} + e^{\lambda B} A' \right|_{R=R_{max}} = 0, \qquad
\left. \dot{C} + e^{\lambda B} C' \right|_{R=R_{max}} = 0.
\label{Eq:Sommerfeld}
\end{equation}
These two boundary conditions set the incoming characteristic fields
corresponding to the evolution equations for $A$ and $C$ to zero, and
thus do absorb high-frequency outgoing waves. However, these boundary
conditions do yield reflections for low-frequency waves propagating
towards large $R$. In our simulations below, we choose $R_{max}$ to be
large enough so that such reflections do not influence the region
where physics is extracted.

\subsection{Numerical implementation}

In order to numerically implement the evolution system
(\ref{Eq:A},\ref{Eq:B},\ref{Eq:C}) we find it convenient to rewrite it
as a first order symmetric hyperbolic system by introducing the
variables $\pi_A := \dot{A}$, $\pi_B := \dot{B}$, $\pi_C := \dot{C}$
and $\psi_A := A'$, $\psi_B := B'$, $\psi_C := C'$. The resulting
system has the form
\begin{eqnarray}
\dot{u} &=& \pi, 
\label{Eq:FOSH1}\\
\dot{\pi} &=& e^{2\lambda B} E^{-1} \frac{d}{dR} ( E\psi) + F(u,\pi,\psi; R),
\label{Eq:FOSH2}\\
\dot{\psi} &=& \frac{d}{dR}\pi,
\label{Eq:FOSH3}
\end{eqnarray}
where $u = (A,B,C)$, $\pi = (\pi_A,\pi_B,\pi_C)$, $\psi =
(\psi_A,\psi_B,\psi_C)$, $E = \mbox{diag}(R^\lambda e^{2\lambda
B},R^{\lambda+2} e^{(\lambda+1)B},R e^{(\lambda +1)B})$ and
$F(u,\pi,\psi; R)$ is a nonlinear term that can be read off from
Eqs. (\ref{Eq:A},\ref{Eq:B},\ref{Eq:C}). This system is then
discretized by the method of lines. Let us start with the description
of our spatial discretization.

We consider first a uniform grid $R_j = j\Delta R$, $j=0,1,2,...N$,
where $\Delta R > 0$ is a fixed mesh size, and replace the functions
$u$, $\pi$ and $\psi$ with gridfunctions $u_j$, $\pi_j$ and $\psi_j$,
$j=0,1,2,...N$, respectively. The evolution system
(\ref{Eq:FOSH1},\ref{Eq:FOSH2},\ref{Eq:FOSH3}) is then approximated by
the semi-discrete system
\begin{eqnarray}
\dot{u}_j &=& \pi_j\; ,\\
\dot{\pi}_j &=& e^{2\lambda B_j} E_j^{-1} D ( E\psi)_j 
 + F(u,\pi,\psi; R_j),\qquad
j=1,2,...N,\\
\dot{\psi}_j &=& (D\pi)_j\; ,
\end{eqnarray}
where $D$ denotes the finite differencing operator
\begin{displaymath}
(D\pi)_j = \left\{ \begin{array}{ll}
 \frac{\pi_{j+1} - \pi_{j-1}}{2\Delta R}\; , & j = 1,2,...N-1, \\
 \frac{\pi_{j} - \pi_{j-1}}{\Delta R}\; ,    & j = N,
\end{array} \right.
\end{displaymath}
which is second order accurate in the interior points $j=1,2,...,N-1$
and first order accurate at the exterior point $j=N$. At the inner
point $j=0$, we use the regularity condition $\psi(t,0) = 0$ and
assume smoothness of $\psi$, which imply
\begin{displaymath}
\lim\limits_{R\to 0} E^{-1} \frac{d}{dR} (E\psi) 
 = \Lambda\, \frac{d\psi}{dR}\; , \qquad
\Lambda = \mbox{diag}(\lambda+1,\lambda+3,2)
\end{displaymath}
by de L'H\^opital's rule. We then evolve $u_0$, $\pi_0$ and $\psi_0$
according to
\begin{eqnarray}
\dot{u}_0 &=& \pi_0\; , \\
\dot{\pi}_0 &=& e^{2\lambda B_0} \Lambda \frac{\psi_1}{\Delta R}
 + F(u,\pi,\psi; 0),\\
\dot{\psi}_0 &=& 0.
\end{eqnarray}
In order to impose the outer boundary conditions
(\ref{Eq:CPBC},\ref{Eq:Sommerfeld}) we replace $(\dot{\psi}_B)_N$ by
\begin{displaymath}
(\dot{\psi}_B)_N 
 = \left[ \pi_A\psi_B + \pi_B( \psi_A - (\lambda+1)\psi_B ) - 6\pi_C\psi_C
 + \frac{1}{R}\left( \pi_A - (\lambda+1)\pi_B \right)
 + \frac{2R}{r_0^2+R^2}\left( \pi_A - \lambda\pi_B - 3\pi_C \right)
 \right]_N\;
\end{displaymath}
and apply Olsson's projection method \cite{pO95} in order to
incorporate the Sommerfeld-type conditions (\ref{Eq:Sommerfeld}). This
method consists in applying the projector
\begin{displaymath}
P = \frac{1}{2} \left( \begin{array}{cc} 
  1              & -e^{\lambda B} \\
 -e^{-\lambda B} &  1
\end{array} \right)
\end{displaymath}
to the two-vectors $(\dot{\pi}_A,\dot{\psi}_A)_N$ and
$(\dot{\pi}_C,\dot{\psi}_C)_N$. The projected fields then satisfy
\begin{displaymath}
(\dot{\pi}_A)_N + e^{\lambda B_N}(\dot{\psi}_A)_N = 0, \qquad
(\dot{\pi}_C)_N + e^{\lambda B_N}(\dot{\psi}_C)_N = 0,
\end{displaymath}
which agrees with (\ref{Eq:Sommerfeld}) in the high-frequency
approximation.

In our simulations, we choose units in which $r_0 = 1$. In order to
improve the accuracy of our code, we replace the radial coordinate $R$
by a computational coordinate $x$ with
\begin{equation}
R = \frac{x}{1 - k\frac{x}{x_{max}}}\; ,\qquad
0 \leq x \leq x_{max}\; ,
\end{equation}
where $0 \leq k < 1$ is a ``stretching'' parameter. A uniform grid
$x_j = j\Delta x$, $j=0,1,2,...N$ in $x$ is then used, where $\Delta x
= x_{max}/N$. Since $dR = dx/(1 - k x/x_{max})^2$, a uniform grid in
$x$ implies (for $k > 0$) a non-uniform grid for $R$, with a smaller
mesh size near the bubble than near the outer boundary. Since the
fields have their largest gradient near the bubble, this
non-uniformity in $R$ helps better resolve the dynamics. Typically, we
use the values $x_{max} = 10$ and $k = 0.8$ in our simulations which
implies that $R_{max}$ is fifty times as large as $r_0$.

Next, we use a third order Runge-Kutta algorithm in order to perform
the time integration. In our simulations, we use $0.2$ as a value for
the Courant-Friedrichs-Levy factor $\Delta t/\Delta x$. In
\cite{dNlLoSmT06} we have shown for a related model problem consisting
of the reduced wave equation for spherically symmetric solutions that
this discretization leads to a stable and convergent scheme.

Finally, initial data for $u$, $\pi$ and $\psi$ is obtained by
comparing the metrics (\ref{Eq:PrototypeMetricNewCoords}) and
(\ref{Eq:MetricEvolution}). This yields
\begin{displaymath}
A = a + \lambda\, b, \qquad
B = b, \qquad
C = 0, \qquad
\psi_A = \frac{da}{dR} + \lambda\frac{db}{dR}\; , \qquad
\psi_B = \frac{db}{dR}\; ,\qquad
\psi_C = 0,
\end{displaymath}
and $\pi_A = \pi_B = \pi_C = 0$, where
\begin{displaymath}
a = -\frac{1}{2}\log(W/R^2) - \frac{1}{2}\log(r_0^2 + R^2),
\qquad
b = \frac{1}{2}\log(H/R^2) + \frac{1}{2}\log(r_0^2 + R^2),
\end{displaymath}
and where we recall that $W/R^2$ and $H/R^2$ converge to a finite
value as $R\to 0$.

%%%%%%%%%%%%%%%%%%%%%%%%%%%%%%%%%%%%%%%%%%%%%%%%%%%%%%%%%%%%%%
\section{Results}
\label{Sect:Results}
%%%%%%%%%%%%%%%%%%%%%%%%%%%%%%%%%%%%%%%%%%%%%%%%%%%%%%%%%%%%%%

The numerical evolution of the initial data discussed in section
\ref{Sect:Idata} suggest that the bubble eventually collapses to the
Schwarzschild-Tangherlini solution, regardless whether or not the
bubble is initially expanding or collapsing. This is shown in detail
in the following subsections.

\subsection{Evolution of the bubble area \label{bubbleareas}}

\begin{figure}[htp]
\includegraphics[width=8cm]{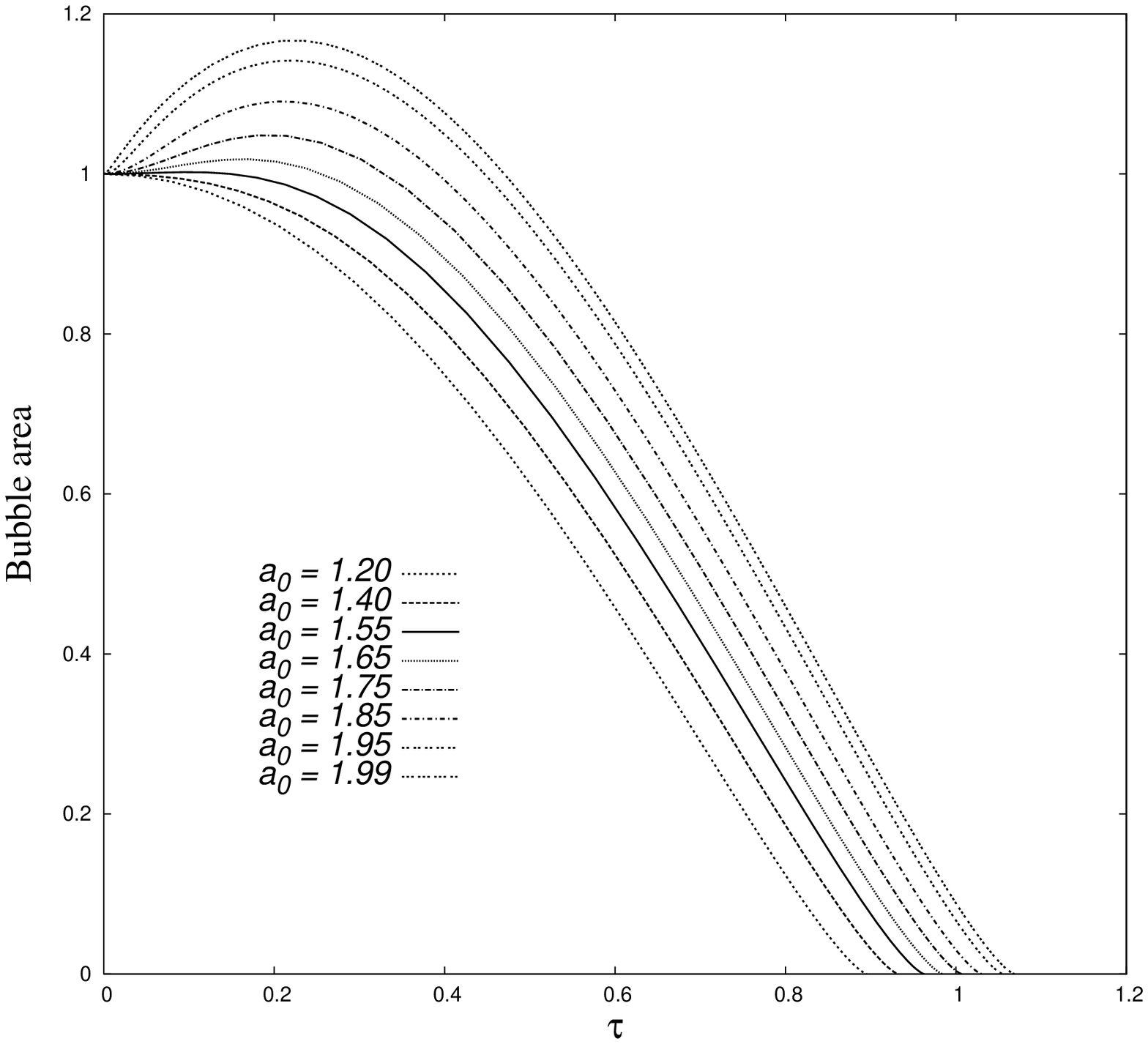}
\includegraphics[width=8cm]{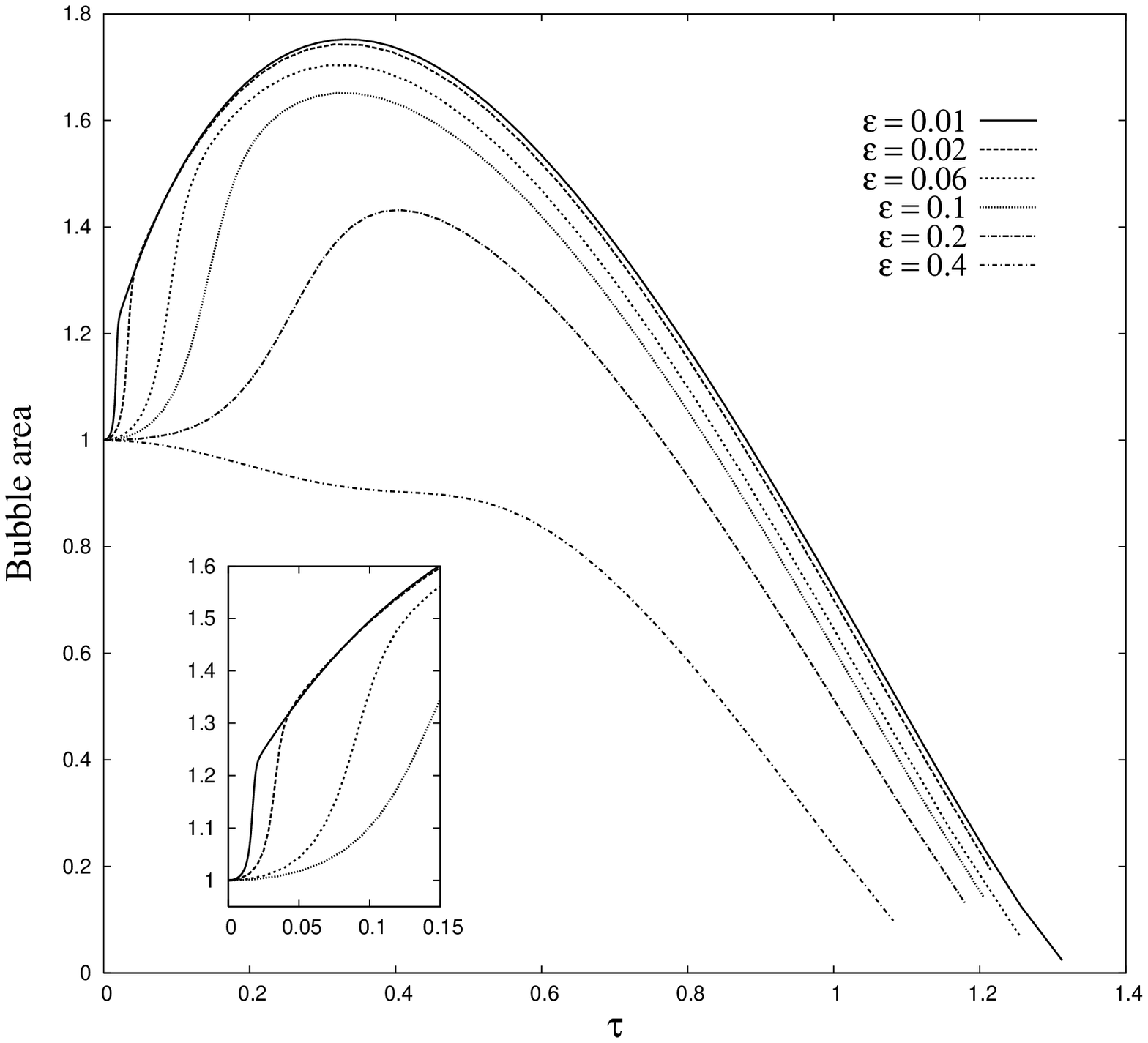}
\caption{\label{Fig:bubble_area} We show the bubble area (normalized
by its initial area) versus proper time evaluated at the bubble. Left:
evolution of initial data of type 2 with several choices for the
initial parameter $a_0$. Right: evolution of type 4 initial data for
$c_2 = 1$ and different values of the parameter $\varepsilon$. The
inset shows an initial violent expansion of the bubble area.
Nevertheless, this behavior changes and the bubble starts
decelerating. In all cases the area shrinks to zero after a finite
proper time.}
\end{figure}

In Fig. \ref{Fig:bubble_area} we show the typical dynamical behavior
of the bubble for type 2 and 4 initial data. More specifically, we
show the bubble's area as a function of proper time at the bubble. It
can be seen that even for type 4 initial data configurations with a
violent initial expansion the bubble finally collapses in a finite
proper time. We also explored a wider range of values for the
parameters $a_0$ and $\varepsilon$, also evolved type 3 initial data
and found that the result is the same in all cases: the bubble area
shrinks to zero after a finite proper time.

The numerical parameters used in these calculations are: $R_{max}=50$,
$k=0.8$, $N=1000$. In the case of the type 2 data, we use the gauge
condition (\ref{Eq:GaugeCondition}) with $\lambda=1$. Because of the
singularity avoidance property of this gauge condition, we are able to
continue the evolution for some time after the bubble area has
shrinked to zero and are able to follow the evolution for long enough
time to observe the formation of an apparent horizon and its settling
down to an equilibrium black hole (see next subsection).  However, for
the evolution of type 4 data we observe the development of large
gradients in the derivatives of the fields, similar to those found in
a shock formation situation. This happens before the turning point in
the evolution of the bubble area and we are unable to continue the
evolution. For this reason we choose the gauge condition
(\ref{Eq:GaugeCondition}) with $\lambda=0$ instead. As mentioned above
the evolution equations are semilinear in this case, and thus there
cannot be shocks due to crossing of characteristics.

\subsection{Formation of an apparent horizon}

In order to provide evidence for the bubble settling down to the
Schwarzschild-Tangherlini solution, we first check whether an apparent
horizon forms. This is done by calculating the expansion of the
outgoing null normal defined as $l^a = n^a + s^a$ with $n^a$ the
future-directed timelike normal to the $t=const$ hypersurface and
$s^a$ the outward pointing unit normal to the surfaces $R=const$ which
is orthogonal to $n^a$. For the $t=const$ surfaces of the metric
(\ref{Eq:MetricGeneral}) the expansion reduces to
\begin{displaymath}
\theta = e^{-a} \left ( b'+2c' \right ) + e^{-d} (\dot b + 2 \dot c).
\end{displaymath}

\noindent In terms of our main evolution variables (i.e. after the
rescaling of the lower case variables, and then the redefinition in
terms of the upper case ones), this can be reexpressed as
\begin{displaymath}
e^{A-2C} \theta =
 e^{\lambda B} \left ( B' + \frac{3 R^2 +r_0}{R(R^2+r_0)} \right ) + \dot B.
\end{displaymath}

\noindent The apparent horizon is determined by the outermost surface
defined by $\theta=0$. Since the factor $e^{A-2C}$ is positive, it can
be ignored for our purposes.

After detecting an apparent horizon we evaluate the curvature
invariants $I_N$, $J_N$ and $K_N$ defined in appendix \ref{App:Tansol}
where we replace $r_{EH}$ by the apparent horizon radius $r_{AH}$
defined as the areal radius of the $t=const$ cross sections of the
apparent horizon,
\begin{equation}
r_{AH} := \left[ R(r_0^2 + R^2) e^B \right]^{1/3}.
\label{Eq:rAH}
\end{equation}
In order to check that the horizon is $SO(4)$ symmetric, we also
compute the quantity
\begin{displaymath}
\mu := \frac{ e^{2b} - e^{2c} }{ e^{2b} + e^{2c} }
     = \frac{ R^2 e^{2(B-2C)} - (R^2 + r_0^2) e^{2C} }
            { R^2 e^{2(B-2C)} + (R^2 + r_0^2) e^{2C} }\; ,
\end{displaymath}
on the apparent horizon. This quantity measures how close a given
three-surface $t=const$, $R=const$, is of being a metric $S^3$. In
particular, $\mu=0$ if this surface is a metric (unsquashed) $S^3$.

\begin{figure}[htp]
\includegraphics[width=8cm]{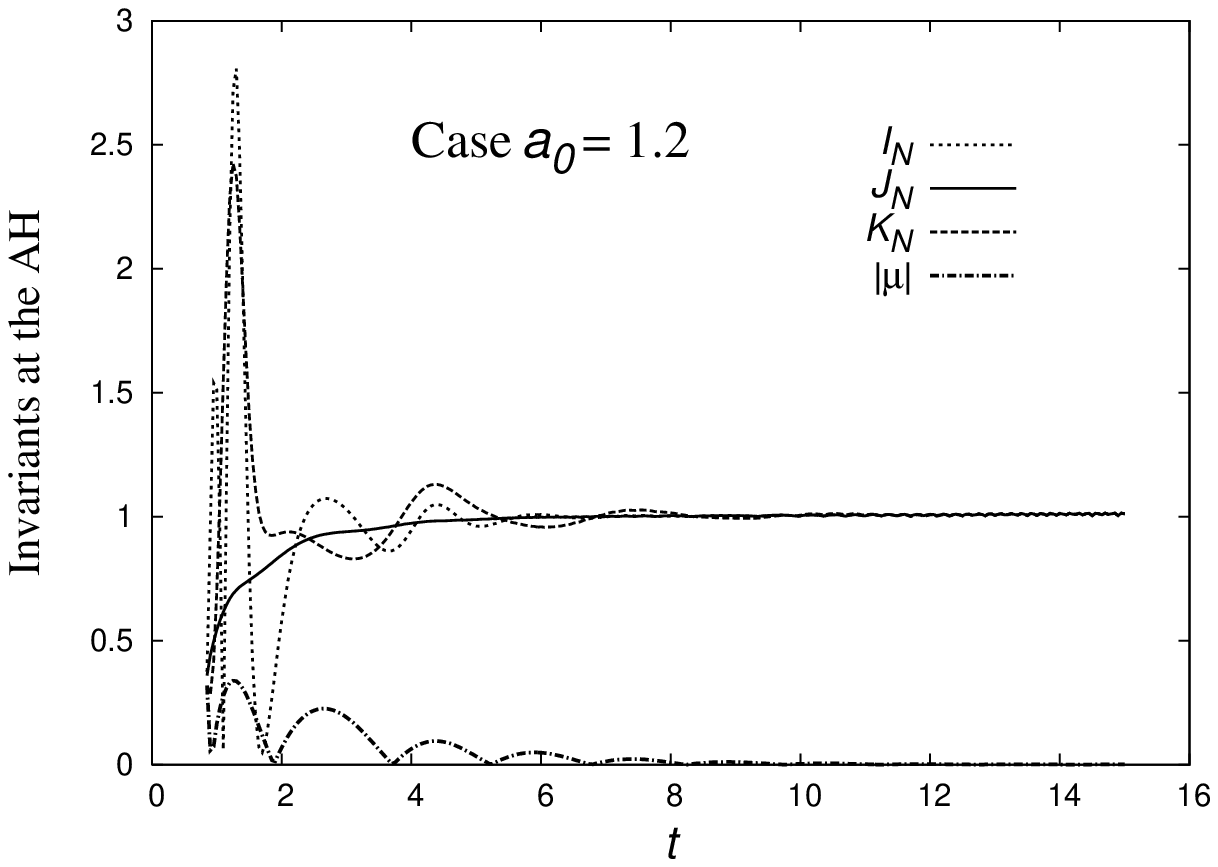}
\includegraphics[width=8cm]{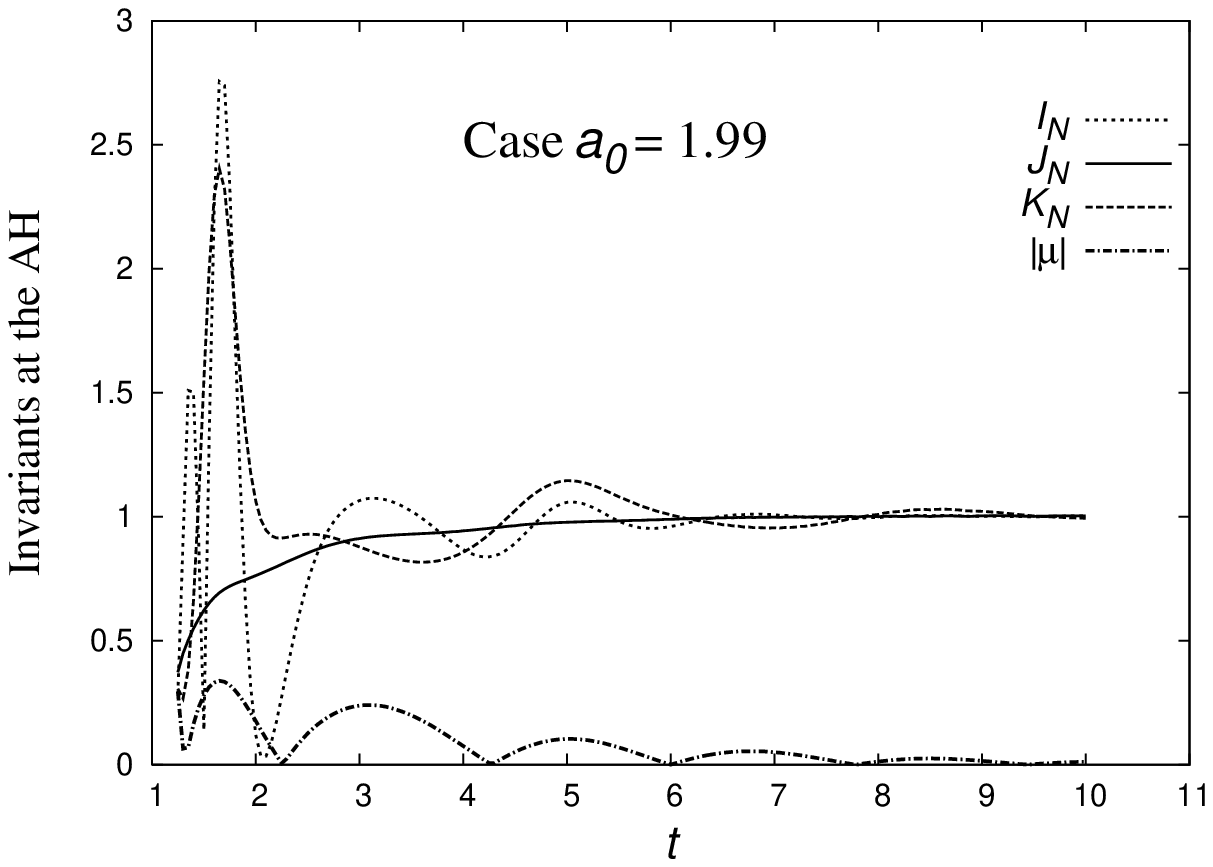}
\caption{\label{Fig:Invariants} The time evolution of the invariants
$I_N$, $J_N$, $K_N$ and $\mu$ for type 2 initial data with two
different values of $a_0$.  The apparent horizon forms at a time
between $t=0.8$ and $0.85$ in the case with $a_0=1.2$ and at a time
between $t=1.25$ and $1.3$ for $a_0=1.99$.}
\end{figure}

The time evolution of the invariants $I_N$, $J_N$, $K_N$ and $\mu$ are
shown in Fig. \ref{Fig:Invariants} for two different values of the
parameter $a_0$ for type 2 initial data. It can be seen that $I_N$,
$J_N$ and $K_N$ converge to $1$ and that $\mu$ converges to zero,
which is a strong indication that the apparent horizon settles down to
the event horizon of the Schwarzschild-Tangherlini solution. The
numerical parameters used in these calculations are: $R_{max}=50$,
$k=0.8$, $N=4000$, $\lambda=1$. For type 4 initial data, where we need
to use the gauge choice (\ref{Eq:GaugeCondition}) with $\lambda=0$ in
order to get to the collapse, we are able to detect the formation of
an apparent horizon, but after that we soon have to stop the
simulation because the resulting gauge condition is not singularity
avoiding. In order to track the late time behavior of the invariants
we switch from $\lambda=0$ to $\lambda=1$ shortly after the bubble
starts collapsing. We then observe the same qualitative features as
for type 2 initial data: an apparent horizon forms and the curvature
invariants $I_N$, $J_N$ and $K_N$ converge to one and $\mu$ to
zero. Therefore, it seems that in this case too the solution settles
down to a Schwarzschild-Tangherlini black hole.

\begin{figure}[htp]
\includegraphics[width=8cm]{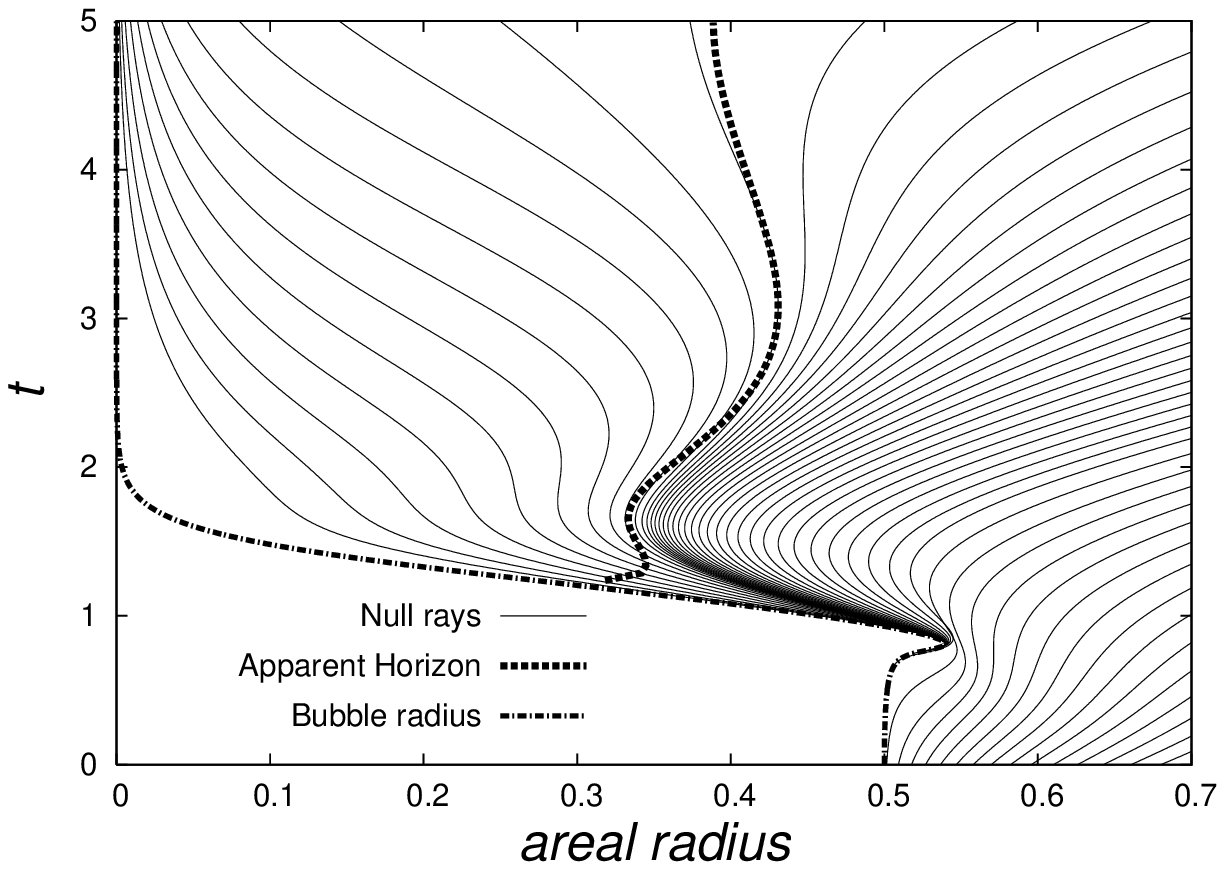}
\includegraphics[width=8cm]{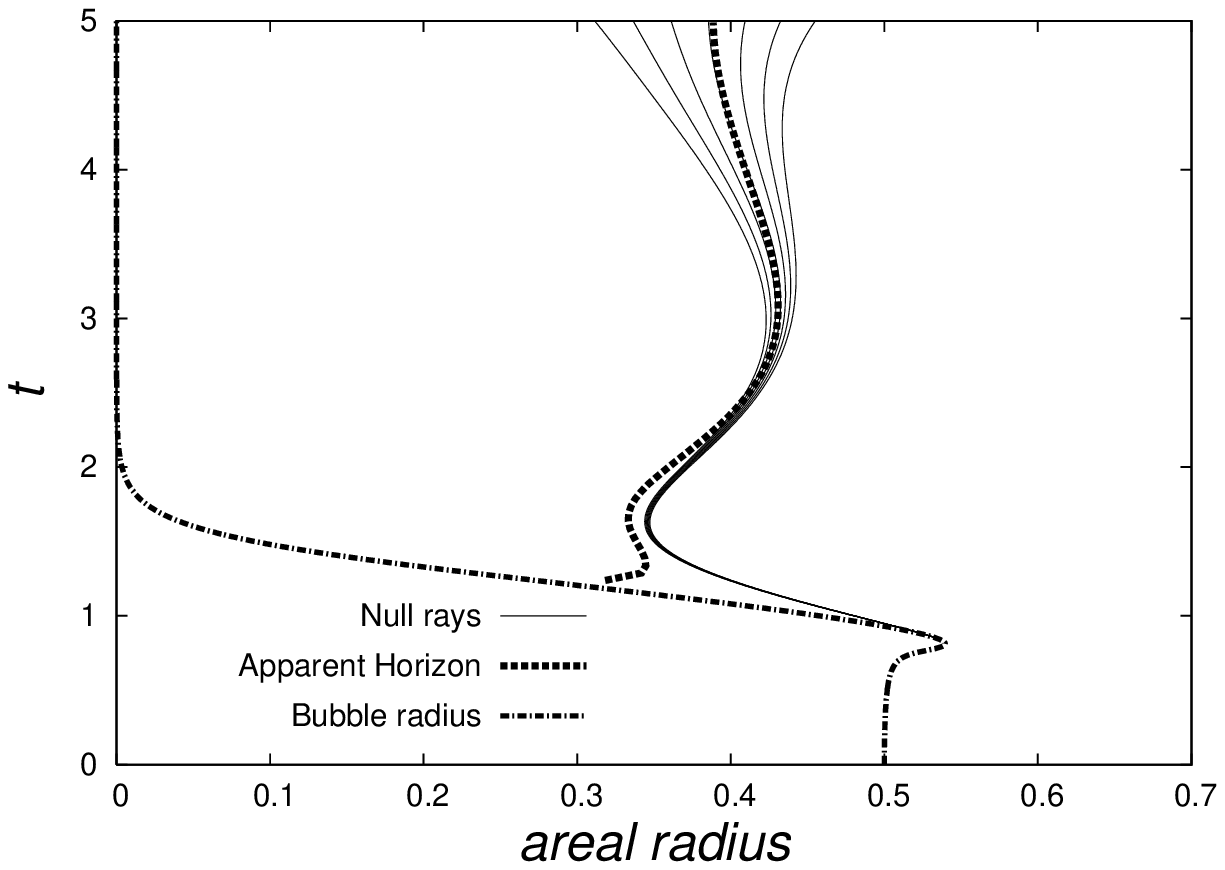}
\caption{\label{Fig:NullRays} Left: the past-directed ingoing null
rays for the numerical spacetime obtained from type 2 initial data
with $a_0=1.99$. The dotted line corresponds to the location of the
apparent horizon and the dotted-dashed line indicates the location of
the bubble area. Right: only a small subset of rays near the location
of the apparent horizon is shown. This indicates that an event horizon
forms at the bubble shortly after the bubble reaches its maximum size.
As can be seen, the apparent horizon at late times $t \gtrsim 2.5$ is
a good approximation for the event horizon. Here, time $t$ refers to
coordinate time and not proper time as in Fig. \ref{Fig:bubble_area}
and the areal radius refers to the geometric radius of the two-spheres
which is equal to $e^{C}\sqrt{R^2 + r_0^2}/2$. Although not apparent
from the plot, we have verified that the area of the $S^3$ cross
sections of the event horizon does grow in time, as expected from the
area theorem
\cite{pCeDgGrH01}.}
\end{figure}

As indicated, all cases explored where the bubbles begin expanding
turn around and eventually collapse. This collapse is hidden however
from external observers by the appearance of an event horizon before
the bubble shrinks to zero size.  An important issue to examine is the
size of the bubble when a horizon forms as if it were significantly
smaller than the bubble's initial size string effects would become
relevant and the present classical analysis would not apply. To
examine this behavior we consider the type 2 data family of initially
expanding bubbles and obtain the size of the bubble when an apparent
horizon first appears. The results (shown in table \ref{Tab:AH})
indicate that the bubble only shrinks to about $1/3$ of its initial
size by the time the horizon forms. Therefore, a ``classically-sized''
bubble would seem to form a horizon while itself is still in the
classical regime as suggested by its apparent horizon size. Certainly
one should be careful with this argument since the apparent horizons
are slicing-dependent. However, by monitoring the behavior of null
rays one can also study the formation of an event horizon and the
overall causal structure of the resulting spacetime. To this end we
trace past-directed ingoing null rays from a late-time slice of our
simulation. This is shown in Fig. \ref{Fig:NullRays} for the type 2
initial data with $a_0=1.99$. As seen from the plot, the event horizon
does not intersect the initial hypersurface but branches off the
bubble shortly after the the bubble reaches its maximum size.  We
would like to mention that while the results discussed above, and the
ones included in the table illustrate what is observed for the case of
type 2 initial data, a similar behavior is obtained for the other data
types discussed.

\begin{table}[h]
\center
\begin{tabular}[c]{|c||l|l|l|l|l|l|}\hline
$a_0$ & 1.55 & 1.65 & 1.75 & 1.85 & 1.95 & 1.99 \\ 
\hline
Bubble size & 0.27 & 0.27 & 0.28 & 0.29 & 0.29 & 0.29 \\
\hline
\end{tabular}
\caption{Size of the bubble when the apparent horizon first appears
for different values of the initial data parameter $a_0$.}
\label{Tab:AH}
\end{table}

\begin{figure}[htp]
\includegraphics[width=8cm]{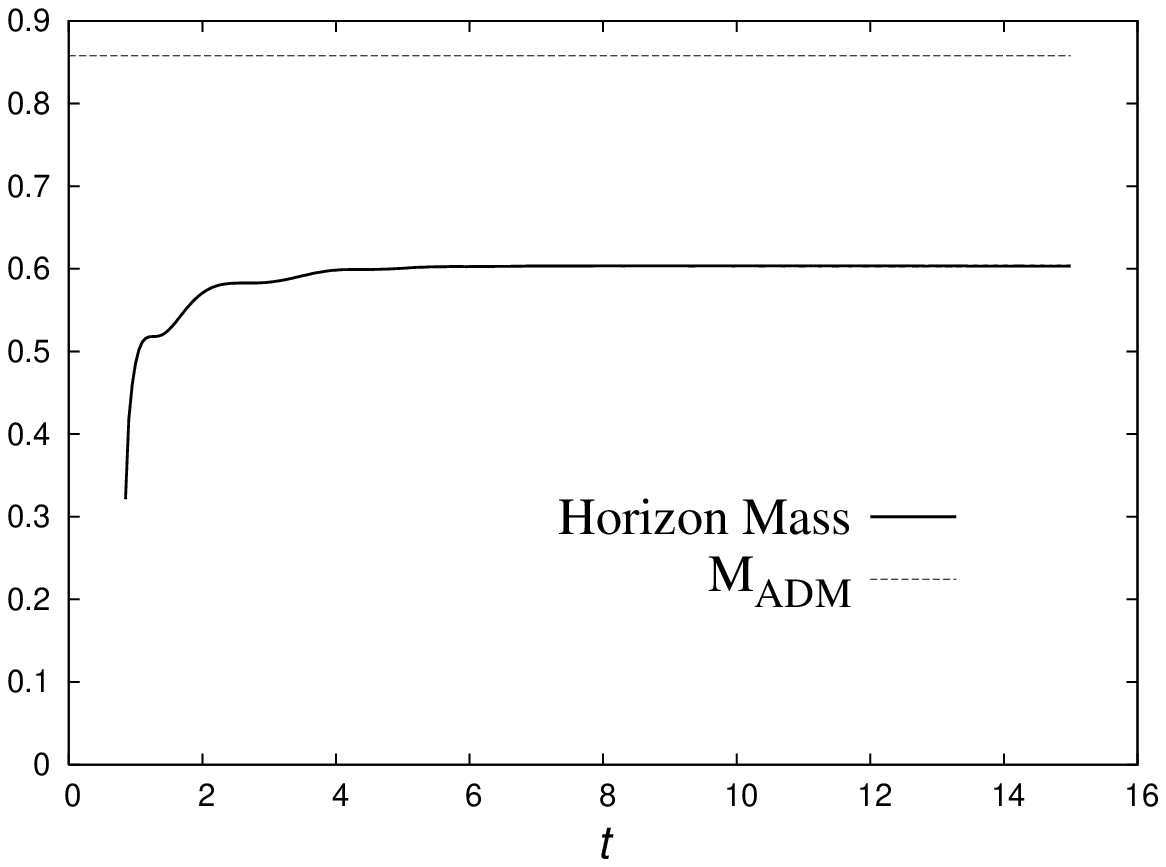}
\includegraphics[width=8cm]{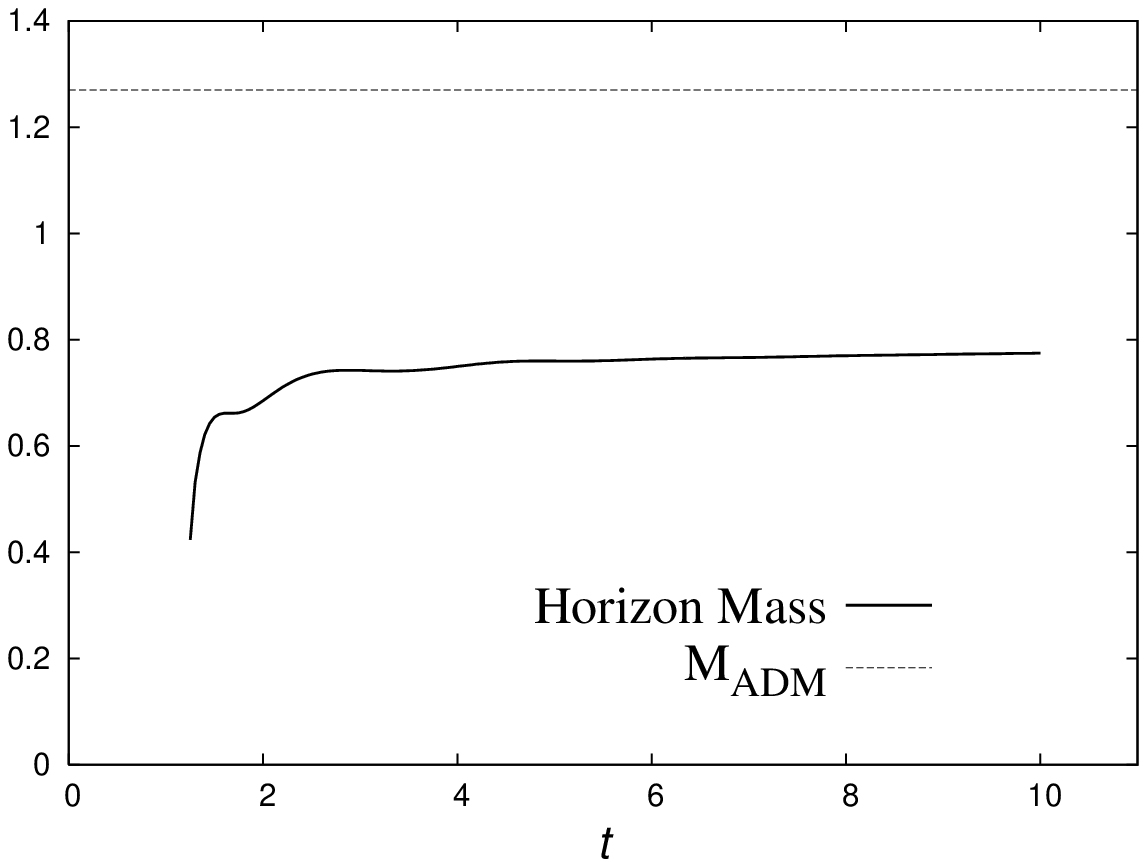}
\caption{\label{Fig:Masses} The $M_{ADM}$ and horizon masses vs time 
are shown for the cases $a_0=1.2$ (left) and $a_0=1.99$ (right).}
\end{figure}

Further insight of the spacetime behavior can be obtained by comparing
the ADM mass with the irreducible mass of the formed horizon. In
Fig. \ref{Fig:Masses} we compare the ADM mass with the apparent
horizon mass of the final black hole which is defined as
$M_{AH}=\frac{3\pi}{8} r_{AH}^{2}$ where $r_{AH}$ is the apparent
horizon radius defined in Eq. (\ref{Eq:rAH}). The ADM mass is
determined from the parameter $a_0$ in the initial data
\cite{kC07b}. The difference between the two masses indicates that
significant radiation is produced during the collapse of the bubble.
The numerical parameters used in these calculations are: $R_{max}=50$,
$k=0.8$, $N=4000$, $\lambda=1$.

\subsection{Convergence checks}

\begin{figure}[htp]
\includegraphics[width=8cm]{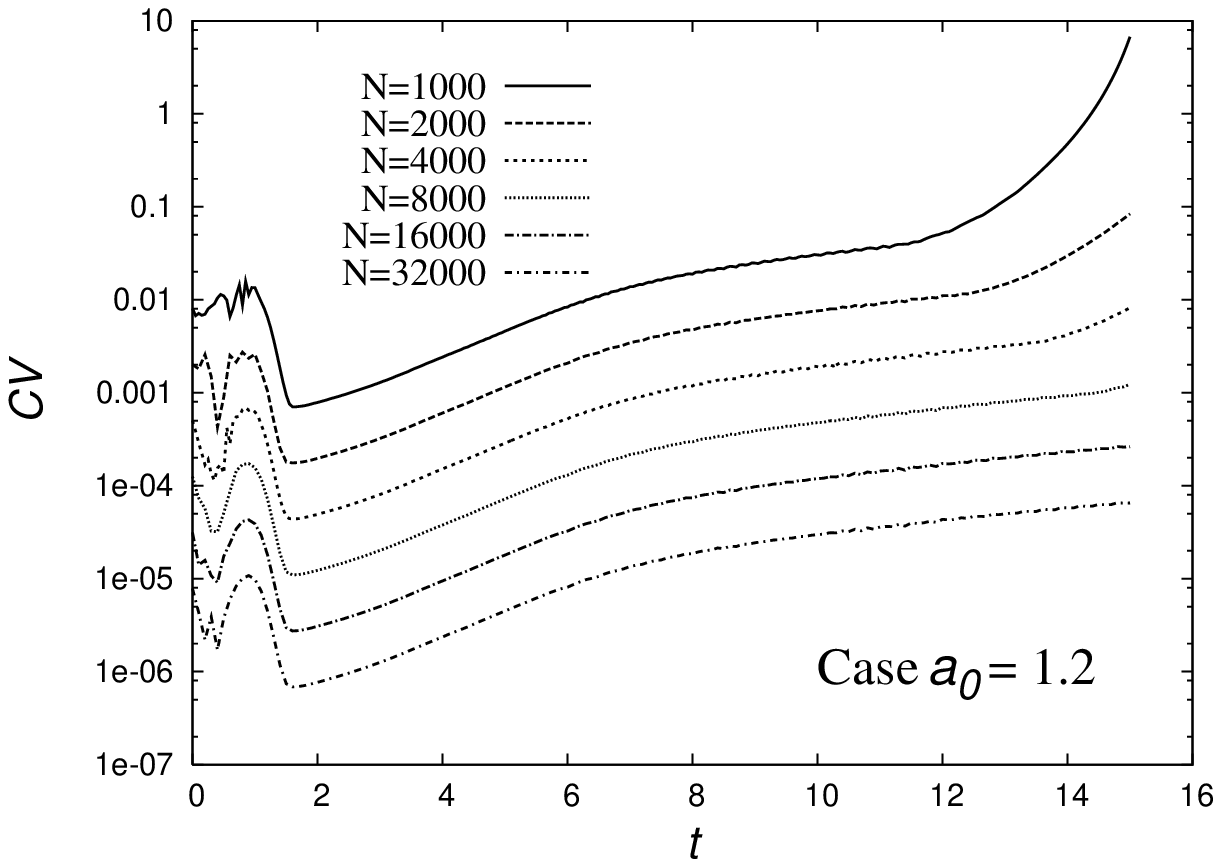}
\includegraphics[width=8cm]{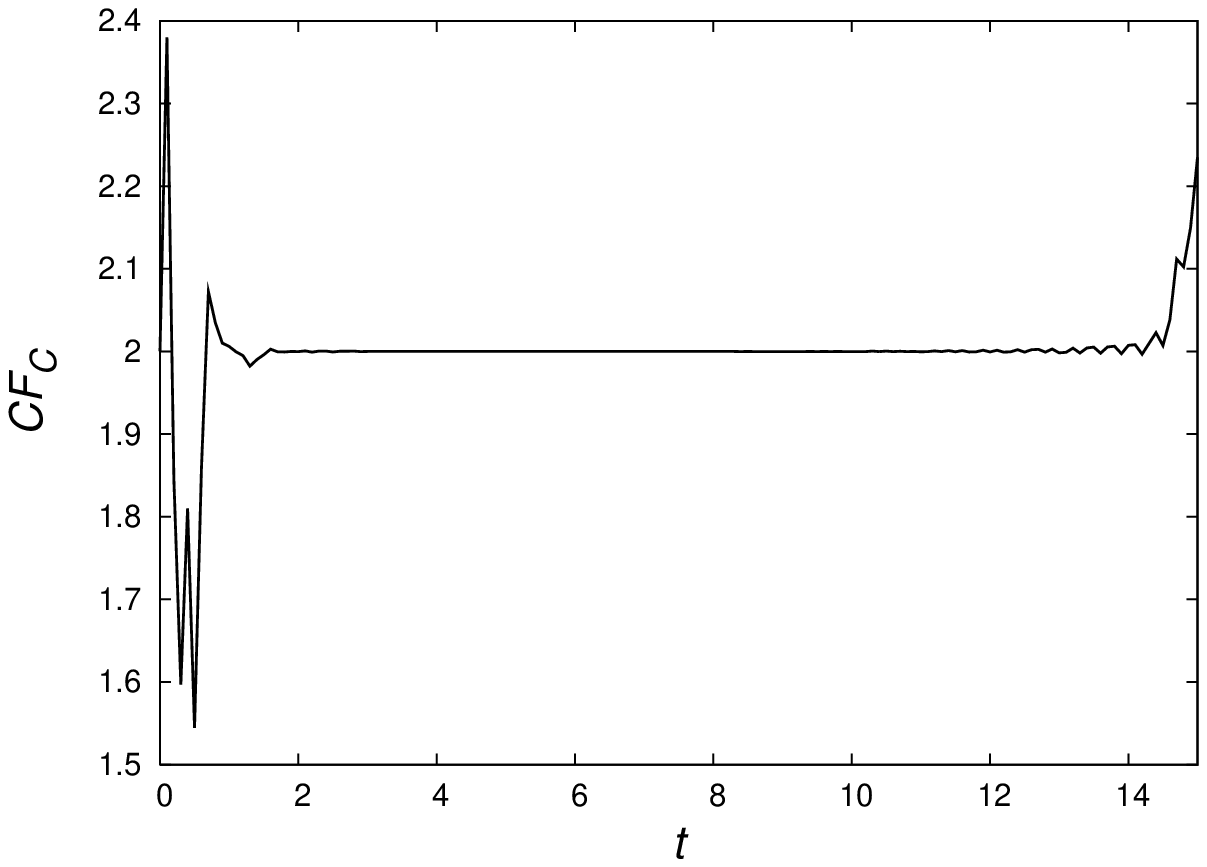}
\caption{\label{Fig:HamMomConvergence1.2} Left: the quantity $CV =
\sqrt{ H_\infty^2 + M_\infty^2}$ vs time for various resolutions, for
the value $a_0=1.2$. Right: the convergence factor $CF_C$ vs
time. This factor is defined as $CF_C:=\log_2( CV(N=8000)/CV(N=16000)
)$.}
\end{figure}

\begin{figure}[htp]
\includegraphics[width=8cm]{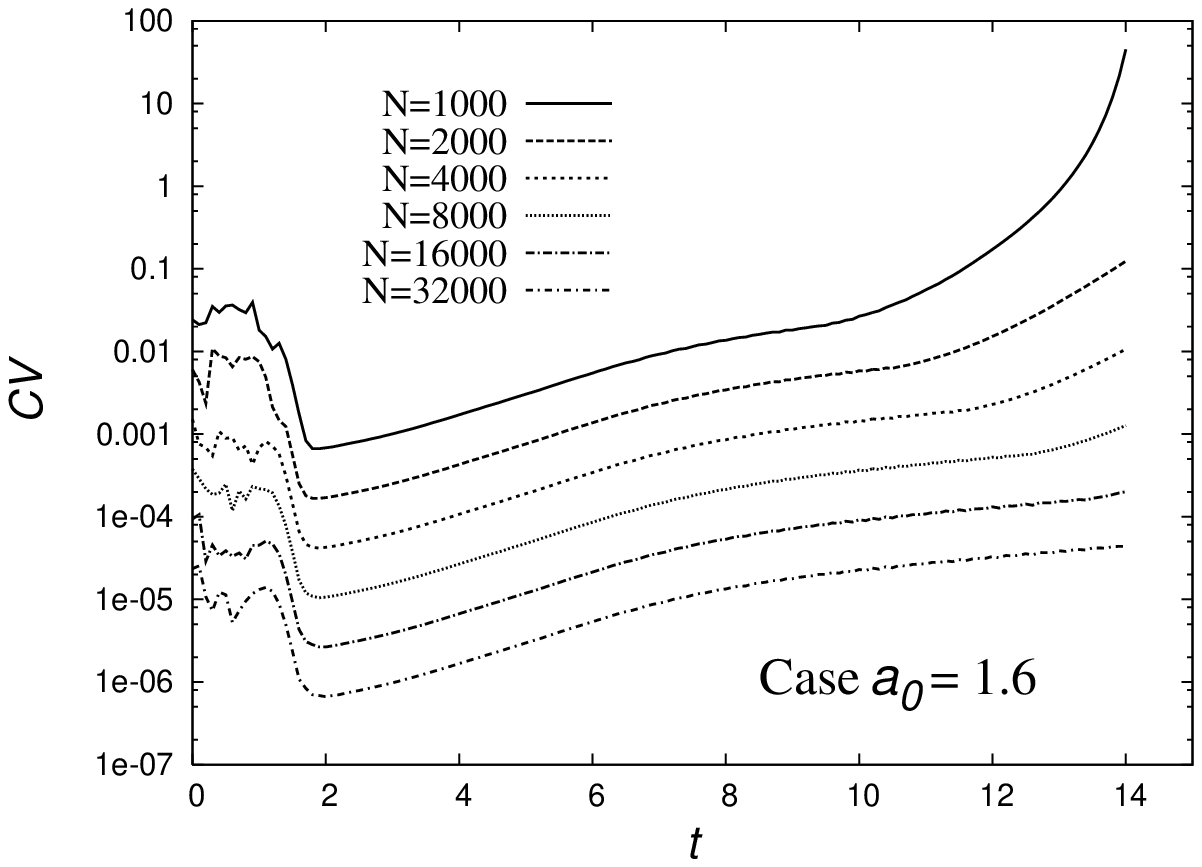}
\includegraphics[width=8cm]{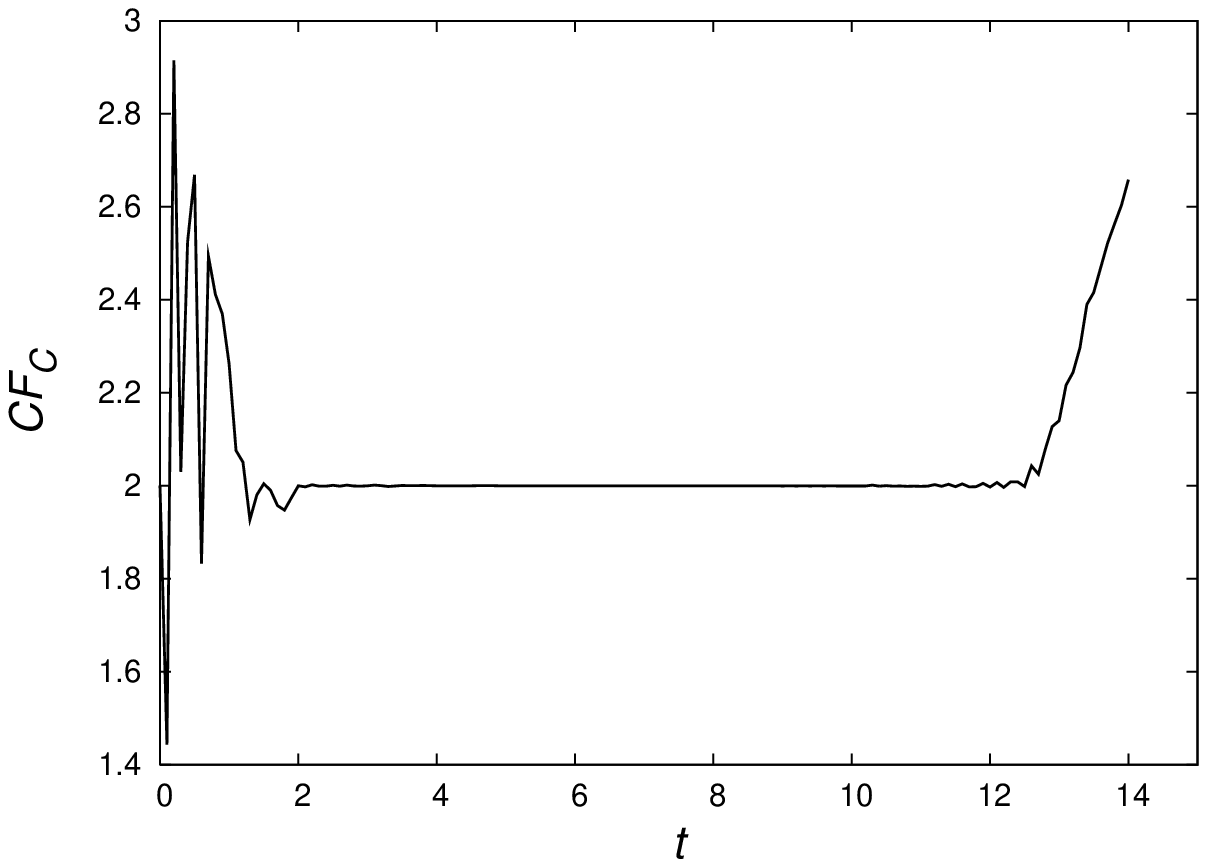}
\caption{\label{Fig:HamMomConvergence1.6} The same as in the previous
figure, but for the value $a_0=1.6$. }
\end{figure}

In order to verify the convergence of our numerical approximation we
performed several tests. First, we verify that the constraint errors
become smaller as resolution is increased. This is shown in Figs.
\ref{Fig:HamMomConvergence1.2} and \ref{Fig:HamMomConvergence1.6} for
type 2 initial data with two different values of $a_0$. In these
figures, we compute the quantity $CV = \sqrt{H_\infty^2 + M_\infty^2}$
where
\begin{displaymath}
H_\infty 
 := \max\limits_{1 \leq j \leq N} \frac{R_j}{\sqrt{1 + R_j^2}}{\cal H}_j\; ,
\qquad
M_\infty 
 := \max\limits_{1 \leq j \leq N} \frac{R_j}{\sqrt{1 + R_j^2}}{\cal M}_j\; .
\end{displaymath}
As Figs. \ref{Fig:HamMomConvergence1.2} and
\ref{Fig:HamMomConvergence1.6} indicate, the constraint errors show
second order convergence to zero as resolution is increased. The
numerical parameters used in these calculations are: $R_{max}=50$ and
$k=0.8$.

\begin{figure}[htp]
\includegraphics[width=8cm]{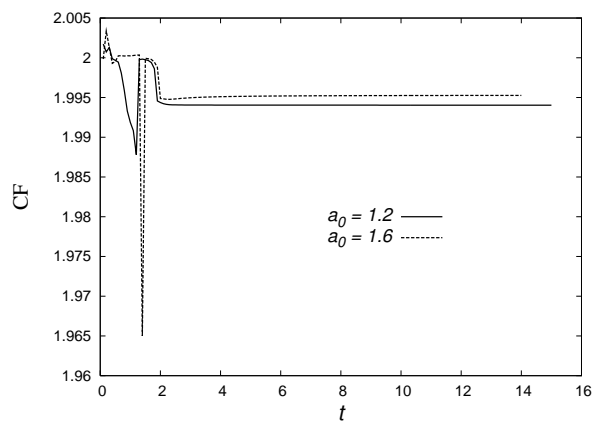}
\caption{\label{Fig:SelfConvergence} Convergence factor $CF$ vs time
for the case $a_0 = 1.6$. This factor is defined as $CF = \log_2( \|
A(N=8000) - A(N=16000) \|_\infty/ \| A(N=16000) - A(N=32000)
\|_\infty)$, where $\| A \|_\infty$ denotes the maximum value of $A$
on the computational grid. The fact that $CF$ is nearly equal two
confirms second order convergence of our code.}
\end{figure}

Next, we also perform a self-convergence test for the field $A$. This
is shown in Fig. \ref{Fig:SelfConvergence} which indicates second
order convergence.

We also find convergence for type 3 and type 4 initial data although
the corresponding convergence factors sometimes drop well below two in
those cases, though remains above 1. This drop is due to the large
gradients appearing in these solutions.

%%%%%%%%%%%%%%%%%%%%%%%%%%%%%%%%%%%%%%%%%%%%%%%%%%%%%%%%%%%%%%
\section{Conclusions}
\label{Sect:Conclusion}
%%%%%%%%%%%%%%%%%%%%%%%%%%%%%%%%%%%%%%%%%%%%%%%%%%%%%%%%%%%%%%

In this work we examined the behavior of families of bubbles of
nothing in five-dimensional asymptotically flat spacetimes. The
simulations carried out reveal that independent of their initial
behavior, these bubbles eventually collapse and form a black hole
horizon around them. The formation of the horizon takes place at a
time when the bubble's size is of the order of its initial
size. Hence, starting with a `classically-sized' bubble, the collapse
and black hole formation occurs well above the string/Planck
length. Thus, at least for the family considered, cosmic censorship is
not violated and the overall process takes place at a classical
level. The singularity that eventually forms in the spacetime is
likely the same as that of the Tangherlini solution, thus its black
hole singularity resolution at the quantum level should apply to this
case as well. The behavior of initially expanding bubbles examined in
this work is in sharp contrast with those with the same initial
behavior in the Kaluza-Klein picture \cite{oSlL04}. There the
expansion continues and increases as time progresses and ultimately
its expansion rate approaches light-speeds. This observation seems to
indicate the presence of a fundamental underlying effect causing the
turn-around in the asymptotically flat case. However, we want to
stress that our analysis has not covered all asymptotically flat
bubble solutions presented by Copsey and that we have not examined the
AdS case. It would be extremely interesting to know whether the
behavior found here is generic.

%%%%%%%%%%%%%%%%%%%%%%%%%%%%%%%%%%%%%%%%%%%%%%%%%%%%%%%%%%%%%%
\section{Acknowledgments}
%%%%%%%%%%%%%%%%%%%%%%%%%%%%%%%%%%%%%%%%%%%%%%%%%%%%%%%%%%%%%%
We would like to thank K. Copsey for many stimulating discussions and
for making available his results prior to publication. We also want to
thank Vijay Balasubramanian, Robert Myers, Jorge Pullin and Thomas
Zannias for useful discussions. Last we wish to thank Jorge Pullin for
bringing this problem to our attention. This work was supported in
part by grants from NSF: PHY-0326311 and PHY-0554793 to Louisiana
State University, by grants CIC: 4.9 and 4.20 to Universidad
Michoacana and PROMEP: UMICH-PTC-121, UMSNH-CA-22 from SEP
Mexico. L.L. thanks CIAR for support and is grateful to the Research
Corporation for financial support. O.S. thanks Louisiana State
University for its hospitality.

\appendix
%%%%%%%%%%%%%%%%%%%%%%%%%%%%%%%%%%%%%%%%%%%%%%%%%%%%%%%%%%%%%%
\section{Expressions for the curvature tensor}
\label{App:Exp}
%%%%%%%%%%%%%%%%%%%%%%%%%%%%%%%%%%%%%%%%%%%%%%%%%%%%%%%%%%%%%%

In this appendix we compute the curvature and Ricci tensors belonging
to the metric
\begin{equation}
ds^2 = -e^{2d} dt^2 + e^{2a} dR^2 
     + e^{2b} \left( dz + \cos\vartheta\, d\varphi \right)^2 
     + e^{2c} d\Omega^2\, ,
\label{Eq:MetricGeneral}
\end{equation}
where $d$, $a$, $b$ and $c$ are functions of $t$ and $R$ only, and
where $d\Omega^2$ denotes the standard metric on $S^2$. In order to do
so, it is convenient to introduce the following orthonormal basis of
one-forms,
\begin{displaymath}
\theta^0 = e^d dt, \qquad
\theta^1 = e^a dR, \qquad
\theta^2 = e^b\left( dz + \cos\vartheta\, d\varphi \right), \qquad
\theta^3 = e^c d\vartheta, \qquad
\theta^4 = e^c \sin\vartheta\, d\varphi.
\end{displaymath}
With respect to this, the curvature two-form $\Omega_{ij} =
\frac{1}{2} R_{ijkl}\theta^k \wedge \theta^l$ is
\begin{eqnarray}
\Omega_{01} &=& -\kappa\; \theta^0 \wedge \theta^1,
\nonumber\\
\Omega_{02} &=& -\beta_0\; \theta^0 \wedge \theta^2 
                -\beta_1\; \theta^1 \wedge \theta^2
 + (\dot{b} - \dot{c}) e^{b-2c-d}\theta^3 \wedge \theta^4,
\nonumber\\
\Omega_{0A} &=& -\gamma_0\; \theta^0 \wedge \theta^A 
                -\gamma_1\; \theta^1 \wedge \theta^A
 + \frac{1}{2}(\dot{b} - \dot{c}) e^{b-2c-d}
    \theta^2 \wedge \varepsilon_{AB}\theta^B,
\nonumber\\
\Omega_{12} &=& -\beta_1\; \theta^0 \wedge \theta^2 
                -\beta_2\; \theta^1 \wedge \theta^2
 + (b' - c') e^{b-2c-a}\theta^3 \wedge \theta^4,
\nonumber\\
\Omega_{1A} &=& -\gamma_1\; \theta^0 \wedge \theta^A 
                -\gamma_2\; \theta^1 \wedge \theta^A
 + \frac{1}{2}(b' - c') e^{b-2c-a}\theta^2 \wedge \varepsilon_{AB}\theta^B,
\nonumber\\
\Omega_{2A} &=& -\frac{1}{2}(\dot{b}-\dot{c}) e^{b-2c-d}
                 \theta^0 \wedge \varepsilon_{AB}\theta^B
                -\frac{1}{2}(b'-c') e^{b-2c-a}
                 \theta^1 \wedge \varepsilon_{AB}\theta^B
 + \delta_1\; \theta^2 \wedge \theta^A,
\nonumber\\
\Omega_{34} &=& (\dot{b}-\dot{c}) e^{b-2c-d}\theta^0 \wedge \theta^2
              + (b'-c') e^{b-2c-a} \theta^1 \wedge \theta^2
              + \delta_2\; \theta^3 \wedge \theta^4.
\nonumber
\end{eqnarray}
Here, $A,B\in \{ 3,4 \}$, $\varepsilon_{34} = -\varepsilon_{43} = 1$,
$\varepsilon_{33} = \varepsilon_{44} = 0$, a prime and a dot denote
differentiation with respect to $R$ and $t$, respectively, and
\begin{eqnarray}
\kappa &=& e^{-2d}\left[ \ddot{a} + \dot{a}(\dot{a}-\dot{d}) \right]
         - e^{-2a}\left[ d'' + d'(d'-a') \right],
\nonumber\\
\beta_0 &=& e^{-2d}\left[ \ddot{b} + \dot{b}(\dot{b}-\dot{d}) \right]
          - e^{-2a} b' d',
\nonumber\\
\beta_1 &=& e^{-a-d}\left[ \dot{b}' + \dot{b}(b'-d') - \dot{a}\, b' \right],
\nonumber\\
\beta_2 &=& e^{-2a}\left[ b'' + b'(b'-a') \right] - e^{-2d} \dot{a}\,\dot{b},
\nonumber\\
\gamma_0 &=& e^{-2d}\left[ \ddot{c} + \dot{c}(\dot{c}-\dot{d}) \right]
          - e^{-2a} c' d',
\nonumber\\
\gamma_1 &=& e^{-a-d}\left[ \dot{c}' + \dot{c}(c'-d') - \dot{a}\, c' \right],
\nonumber\\
\gamma_2 &=& e^{-2a}\left[ c'' + c'(c'-a') \right] - e^{-2d} \dot{a}\,\dot{c},
\nonumber\\
\delta_1 &=& e^{-2d} \dot{b}\,\dot{c} - e^{-2a} b' c'
           + \frac{1}{4}\, e^{2b-4c},
\nonumber\\
\delta_2 &=& e^{-2d} \dot{c}^2 - e^{-2a} c'^2 
           - \frac{3}{4}\, e^{2b-4c} + e^{-2c}.
\nonumber
\end{eqnarray}
The corresponding components of the Ricci tensor are obtained from
$R_{ij} = \Omega^k{}_i(e_k,e_j)$, where $e_j$, $j=0,1,...,4$, is the
F\"unfbein corresponding to the orthonormal basis
$\theta^0,\theta^1,...,\theta^4$ of one-forms. Explicitly, we find
\begin{eqnarray}
R_{11} &=& e^{-2d}\left[ 
    \ddot{a} + \dot{a}\,(\dot{a}+\dot{b}+2\dot{c}-\dot{d}) \right]
 - e^{-2a}\left[ d'' + b'' + 2c'' + d'(d'-a') + b'(b'-a') + 2c'(c'-a') \right],
\\
R_{22} &=& e^{-2d}\left[ 
    \ddot{b} + \dot{b}\,(\dot{a}+\dot{b}+2\dot{c}-\dot{d}) \right]
 - e^{-2a}\left[ b'' + b'(b'+2c'+d'-a') \right] + \frac{1}{2}\, e^{2b-4c},
\\
R_{33} = R_{44} &=& e^{-2d}\left[ 
    \ddot{c} + \dot{c}\,(\dot{a}+\dot{b}+2\dot{c}-\dot{d}) \right]
 - e^{-2a}\left[ c'' + c'(b'+2c'+d'-a') \right] - \frac{1}{2}\, e^{2b-4c}
 + e^{-2c},
\\
R_{01} = R_{10} &=& -e^{-a-d}\left[
 \dot{b}' + 2\dot{c}' - d'(\dot{b} + 2\dot{c}) 
 + b'(\dot{b}-\dot{a}) + 2c'(\dot{c}-\dot{a}) \right],
\\
R_{00} &=& 2G_{00} - R_{11} - R_{22} - R_{33} - R_{44}\; .
\end{eqnarray}
Here, $G_{00}$ refers to the $00$ components of the Einstein tensor,
which is
\begin{equation}
G_{00} = e^{-2d}\left[ \dot{a}(\dot{b} + 2\dot{c}) 
                       + \dot{c}(\dot{c} + 2\dot{b}) \right]
 - e^{-2a}\left[ b'' + 2 c'' + (b'-a')(b'+2c') + 3 c'^2 \right]
 - \frac{1}{4}\, e^{2b - 4c} + e^{-2c}.
\end{equation}
The remaining components of $R_{ij}$ are identically zero.

Using the fact that the metric is $SO(3)$ symmetric and homogeneous in
the $z$ direction, several curvature invariants can be defined. For
this, consider the five-metric $ds^2$ in Eq. (\ref{Eq:MetricGeneral}),
the orbit three-metric
\begin{displaymath}
d\bar{s}^2 =  -e^{2d} dt^2 + e^{2a} dR^2 + e^{2b} dz^2
\end{displaymath}
and the projection thereof on the spaces orthogonal to the Killing
field $\partial_z$
\begin{displaymath}
d\tilde{s}^2 =  -e^{2d} dt^2 + e^{2a} dR^2.
\end{displaymath}
The Kretschmann invariants $I=R_{ijkl}R^{ijkl}$ with respect to these
metrics are
\begin{eqnarray}
\tilde{I} &=& 4\kappa^2,
\\
\bar{I} &=& 4
\left[ 
  \kappa^2 + \beta_{0}^{2} - 2\beta_{1}^{2} + \beta_{2}^{2}
\right],
\\
I &=& 4 
\left[
\kappa^2 + \beta_{0}^{2} - 2\beta_{1}^{2} + \beta_{2}^{2} + 2 \gamma_{0}^{2}
-4\gamma_{1}^{2} + 2\gamma_{2}^{2} + 2\delta_{1}^{2} +\delta_{2}^{2}
-3(\dot{b} - \dot{c})^2 e^{2(b-2c-d)}
+3(b^{\prime} - c^{\prime})^2 e^{2(b-2c-a)}
\right].
\end{eqnarray}

Finally, the twice contracted Bianchi identities $\nabla_i G^i{}_j =
0$ are
\begin{eqnarray}
&& \dot{G}_{00} + (\dot{a} + \dot{b} + 2\dot{c}) G_{00}
 + \dot{a}\, G_{11} + \dot{b}\, G_{22} + \dot{c}( G_{33} + G_{44} )
 = e^{d-a}\left[ G_{01}' + (b' + 2c' + 2d') G_{01} \right],\\
&& \dot{G}_{01} + (2\dot{a} + \dot{b} + 2\dot{c} ) G_{01}
 = e^{d-a}\left[ G_{11}' + d'\, G_{00} + (b' + 2c' + d') G_{11} 
  - b'\, G_{22} - c'( G_{33} + G_{44} ) \right].
\end{eqnarray}

%%%%%%%%%%%%%%%%%%%%%%%%%%%%%%%%%%%%%%%%%%%%%%%%%%%%%%%%%%%%%%
\section{Tangherlini solution and invariants}
\label{App:Tansol}
%%%%%%%%%%%%%%%%%%%%%%%%%%%%%%%%%%%%%%%%%%%%%%%%%%%%%%%%%%%%%%

Tangherlini's solution \cite{fT63} in five dimensions reads
\begin{equation}
ds^2 = -\left (1-\frac{C}{r^2} \right ) dt^2 
      + \left ( 1 -\frac{C}{r^2} \right )^{-1} dr^2 + r^2 d\Omega^2_3
\end{equation}
with $C$ related to the ADM mass by
\begin{equation}
M_{ADM} = \frac{3\pi}{8}\, C.
\end{equation}
For the curvature invariants, we find $\kappa = 3C/r^4$, $\beta_0 =
\beta_2 = \gamma_0 = \gamma_2 = \delta_1 = \delta_2 = C/r^4$, $\beta_1
= \gamma_1 = 0$, and so the curvature invariants $\tilde{I}$,
$\bar{I}$ and $I$ defined in appendix \ref{App:Exp} are
\begin{displaymath}
\tilde{I} = 36\frac{C^2}{r^8}\; , 
\qquad
\bar{I} = 44\frac{C^2}{r^8}\; , 
\qquad
I = 72\frac{C^2}{r^8}\; .
\end{displaymath}
At the event horizon, the following combinations are particularly
simple: $J_N = \tilde{I} r_{EH}^4/36$, $K_N = \bar{I} r_{EH}^4/44$,
$I_N = I r_{EH}^4/72$.  Since $r_{EH}^2 = C$, $I_N = J_N
= K_N = 1$ when evaluated at the horizon.

%%%%%%%%%%%%%%%%%%%%%%%%%%%%%%%%%%%%%%%%%%%%%%%%%%%%%%%%%%%%%%
% Create the reference section using BibTeX:
\bibliography{refs_stringy}
%%%%%%%%%%%%%%%%%%%%%%%%%%%%%%%%%%%%%%%%%%%%%%%%%%%%%%%%%%%%%%
\end{document}